\documentclass[aps,prd,twocolumn,groupedaddress,superscriptaddress]{revtex4-2}

\usepackage{lipsum}
\usepackage{graphicx}  
\usepackage{dcolumn}   
\usepackage{bm}        
\usepackage{amssymb}   
\usepackage{amsmath}
\usepackage{xcolor}
\usepackage{amsmath}
\usepackage{amsfonts}
\usepackage{hyperref}
\usepackage{float}

\usepackage{hyperref}
\hypersetup{colorlinks=true, linkcolor=blue, citecolor=blue, urlcolor=blue}

\usepackage{bm}
\usepackage{xcolor}

\begin{document}
\preprint{APS/123-QED}

\title{Systematic Bayesian investigation of the (2+1)-flavor QCD phase transition in a holographic model}

\author{Liqiang Zhu}
\email{zhuliqiang@scnu.edu.cn}
\affiliation{State Key Laboratory of Nuclear Physics and Technology, Institute of Quantum Matter, South China Normal University, Guangzhou 510006, China}
\affiliation{Guangdong Basic Research Center of Excellence for Structure and Fundamental Interactions of Matter, Guangdong Provincial Key Laboratory of Nuclear Science, Guangzhou 510006, China}

\author{Xun Chen}
\email{chenxun@usc.edu.cn}
\affiliation{School of Nuclear Science and Technology, University of South China, Hengyang 421001, China}
\affiliation{INFN -- Istituto Nazionale di Fisica Nucleare -- Sezione di Bari, Via Orabona 4, 70125 Bari, Italy}

\author{Kai Zhou}
\email{zhoukai@cuhk.edu.cn}
\affiliation{School of Science and Engineering, The Chinese University of Hong Kong, Shenzhen (CUHK-Shenzhen), Guangdong, 518172, China}
\affiliation{School of Artificial Intelligence, The Chinese University of Hong Kong, Shenzhen (CUHK-Shenzhen), Guangdong, 518172, China}

\author{Hanzhong Zhang}
\email{zhanghz@mail.ccnu.edu.cn}
\affiliation{Key Laboratory of Quark and Lepton Physics (MOE) and Institute of Particle Physics, Central China Normal University, Wuhan 430079, China}

\author{Mei Huang}
\email{huangmei@ucas.ac.cn}
\affiliation{School of Nuclear Science and Technology, University of Chinese Academy of Sciences, Beijing 100049, China}

\author{Enke Wang}
\email{wangek@scnu.edu.cn}
\affiliation{State Key Laboratory of Nuclear Physics and Technology, Institute of Quantum Matter, South China Normal University, Guangzhou 510006, China}
\affiliation{Guangdong Basic Research Center of Excellence for Structure and Fundamental Interactions of Matter, Guangdong Provincial Key Laboratory of Nuclear Science, Guangzhou 510006, China}

\begin{abstract}
In this study, we construct a bayesian holographic QCD model by integrating the Einstein-Maxwell-Dilaton (EMD) framework with lattice quantum chromodynamics (QCD) data at zero chemical potential, specifically entropy, the square of the speed of sound, and baryon number susceptibilities, while systematically incorporating error estimates from the lattice QCD (LQCD) results. Leveraging a bayesian inference framework, we first achieve a precise calibration of the model parameters, then perform a comprehensive investigation into the thermodynamic properties of \((2+1)\)-flavor QCD at both zero and finite chemical potentials, and finally provide a prediction for the location of the critical end point (CEP) in the QCD phase diagram. Our results indicate that, under the maximum a posteriori (MAP) estimation, the CEP is located at $(T, \mu)_{\mathrm{MAP}} = (0.119, 0.615)\mathrm{GeV}$. Furthermore, we provide the predicted regions for the CEP at $68\%$ and $95\%$ confidence levels (CL), yielding $(T, \mu)_{68\% \mathrm{CL}} = (0.1176\text{--}0.1204,0.59\text{--}0.63)\mathrm{GeV}$ and $(T, \mu)_{95\% \mathrm{CL}} = (0.1172\text{--}0.1206,0.58\text{--}0.63)\mathrm{GeV}$, respectively. A thorough comparison with predictions from other theoretical models validates the robustness and predictive power of our approach. This work not only establishes a novel analytical framework for holographic modeling but also provides valuable theoretical insights into the phase transitions of strongly interacting matter under extreme conditions.
\end{abstract}
\maketitle

\section{Introduction}
\label{secintro}

Within the framework of the Standard Model, the strong interaction between quarks and gluons is described by Quantum Chromodynamics (QCD). Based on the asymptotic freedom property of QCD, theory predicts that under conditions of sufficiently high temperature or density, the strongly interacting matter will transition from a confined hadronic state to a novel deconfined state, commonly referred to as the quark–gluon plasma (QGP) \cite{BRAHMS:2004adc,PHOBOS:2004zne,STAR:2005gfr,PHENIX:2004vcz, STAR:2010vob, PHENIX:2001hpc, STAR:2003fka}. In this new state, quarks and gluons are no longer restricted to the interiors of hadrons but can interact and propagate freely. This phenomenon reveals the behavior of QCD matter under extreme conditions. The exploration of this transition, and more broadly of the QCD phase diagram, has long been a central focus in high-energy physics and continues to be a vibrant field of investigation \cite{Du:2024wjm,Guenther:2020jwe,Zhou:2023pti,Luo:2020pef, Li:2026hnl, Ma:2023zfj,Fu:2022gou,Zhao:2026mcp,Li:2026lxx,Wang:2026xwa,Xu:2026qvi,Borsanyi:2025dyp,Li:2025lmp,Zhang:2026dny,Mei:2026vay,Liu:2026rmo,Chen:2026mtw}.

The theoretical foundation for studying the QCD phase diagram lies in first-principles lattice QCD (LQCD) calculations. Although the equation of state (EoS) can be reliably determined at finite temperature, zero chemical potential, and small chemical potential, the sign problem inherent in LQCD methods when dealing with finite baryon density remains a significant obstacle. To gain a deeper understanding of the QCD phase diagram, the research community has also developed and employed a range of alternative theoretical frameworks, including the functional renormalization group (FRG) method \cite{Fu:2022gou,Zhang:2017icm}, Dyson–Schwinger equations (DSEs) \cite{Qin:2010nq,Shi:2014zpa,Gao:2016qkh}, and several effective QCD models, such as the random matrix model (RMM) \cite{Halasz:1998qr} and the family of Nambu–Jona-Lasinio (NJL) models \cite{Li:2018ygx,Stephanov:2004wx,Sun:2024anu,Berges:1998rc,Bao:2024glw}. These approaches provide a rich set of theoretical tools and diverse perspectives for understanding the behavior of strongly interacting matter.

In recent decades, the anti–de Sitter/conformal field theory (AdS/CFT) correspondence~\cite{Natsuume:2014sfa,tHooft:1973alw,Gubser:1998bc,Witten:1998qj,Casalderrey-Solana:2011dxg,Erdmenger:2018xqz,Witten:1998zw} has emerged as an important non-perturbative framework for studying hadron physics and QCD matter at high temperatures and densities. This framework not only provides profound insights into the hadron spectrum but also offers new avenues for investigating QCD matter under extreme thermal density and chemical potential conditions. Within this framework, the Einstein–Maxwell–dilaton (EMD) coupled model has demonstrated remarkable efficacy in reproducing several key features of the QGP \cite{Tong:2026dwb,Khan:2026uqc,Bu:2024fhz,Sun:2025uga,Zhang:2025wxi,Zhu:2025xiz,Chen:2025fpd,Deng:2026aht,Wang:2026ump,Yang:2026brr}. Meanwhile, with the advancement of machine learning methods, an increasing number of studies have integrated holographic duality with data-driven strategies. By leveraging experimental data and LQCD data to constrain or invert the bulk spacetime geometry, these efforts aim to construct observation-oriented holographic models. The relevant work encompasses a variety of EMD-based models as well as data-informed methods for determining the bulk metric. The literature is extensive, covering both theoretical studies grounded in AdS/CFT and the broad application of the duality framework, along with attempts to introduce data-driven approaches into holographic modeling \cite{Maldacena:1997re, Erdmenger:2007cm, Brodsky:2014yha, Casalderrey-Solana:2011dxg, Adams:2012th, Li:2016gtz, BitaghsirFadafan:2018uzs, Abt:2019tas, Nakas:2020hyo, Li:2013oda, Gubser:2008ny, DeWolfe:2010he, Li:2013oda, Yang:2015aia, Yang:2014bqa, Dudal:2017max, Dudal:2018ztm, Fang:2015ytf, Liu:2023pbt, Li:2022erd, Critelli:2017oub, Grefa:2021qvt, Arefeva:2021jpa, Arefeva:2020vae, Chen:2018vty, Chen:2017lsf, Chen:2020ath, Zhou:2020ssi, Chen:2019rez, Cao:2022csq, Arcos:2022icf, Cao:2024jgt, Bohra:2020qom, Arefeva:2023jjh, Bohra:2019ebj, Hashimoto:2018ftp, Hashimoto:2018bnb, Akutagawa:2020yeo, Hashimoto:2022eij, Yan:2020wcd, Song:2020agw, Chang:2024ksq, Ahn:2024gjf, Gu:2024lrz, Li:2022zjc, Ahn:2024lkh, Chen:2024ckb, Chen:2024jet, Hashimoto:2026hfz}.

Bayesian inference has been widely applied in heavy-ion physics to address several key analytical problems, such as determining the jet transport coefficient, jet energy loss distributions, estimating the bulk viscosity, and the equation of state of quantum chromodynamics (QCD EoS) \cite{Xie:2022fak, Xing:2023ciw, Wu:2023azi, JETSCAPE:2023ikg, He:2018gks, JETSCAPE:2021rfx, Parkkila:2021tqq, Bernhard:2019bmu, Cheng:2023ucp}. Furthermore, it enables parameter estimation from a set of observables through global fitting techniques \cite{Pratt:2015zsa, Bernhard:2016tnd, Bernhard:2019bmu}. Recently, in Ref. \cite{Hippert:2023bel}, Bayesian inference was employed in the EMD model, where specific functional forms for the dilaton potential $V(\phi)$ and the coupling between the dilaton and the gauge field $f(\phi)$ are assumed. In this study, we adopt the potential reconstruction approach, in which the dilaton field or metric is input to determine $V(\phi)$, and then incorporate Bayesian inference into the EMD model. A notable advantage of the potential reconstruction method in our model is that it yields analytical solutions, thereby providing robust theoretical support for subsequent investigations into various interesting physical problems.

Compared with machine learning–based EMD models, a distinctive feature of this study is the systematic incorporation of LQCD data uncertainties into the parameter inference process. This yields explicit confidence intervals for the derived physical quantities, rather than merely providing best-fit results, thereby enhancing the reliability and robustness of our conclusions.
The remainder of this paper is organized as follows. Section~\ref{sec2} reviews the EMD model, elucidating its foundational principles and its application background in holographic thermal physics. Section~\ref{sec3} describes the Bayesian inference framework for parameter estimation, detailing how LQCD data with errors are integrated into model calibration to achieve more robust inference. Section~\ref{sec4} presents the derivation and analysis of thermodynamic quantities (including energy density, pressure, etc.) at both zero and finite chemical potential through Bayesian inference. Section~\ref{sec5} focuses on the model’s prediction of the CEP location and compares it with results from other theoretical frameworks. Section~\ref{sec6} concludes the paper with a summary of the main findings and an outlook for future research directions.

\section{The review of EMD model}
\label{sec2}

This section presents a systematic overview of the five-dimensional EMD system, grounded in the framework of potential reconstruction \cite{Zhu:2025gxo,Li:2011hp,Cai:2012xh,He:2013qq,Yang:2014bqa,Yang:2015aia,Dudal:2017max,Dudal:2018ztm,Chen:2018vty,Chen:2020ath,Zhou:2020ssi,Chen:2019rez,Lin:2024mct,Zhu:2026dmc,Chen:2025goz}. The action governing this system incorporates three key fields: a gravitational field $g_{\mu \nu}^s$, a Maxwell field $A_\mu$, and a neutral dilaton scalar field. Within the string frame, the system's dynamical evolution is governed by the following equation.
\begin{equation}
\begin{split}
S_b &=\frac{1}{16 \pi G_5} \int d^5 x \sqrt{-g^s} e^{-2 \phi_s}\\
&\quad \times\left[R_s-\frac{f_s\left(\phi_s\right)}{4} F^2+4 \partial_\mu \phi_s \partial^\mu \phi_s-V_s\left(\phi_s\right)\right],
\end{split}
\label{Eq:actionsb}
\end{equation}
In this context, $f(\phi)$ represents the gauge kinetic function coupled to the Maxwell field $A_\mu$, while $V(\phi)$ denotes the dilaton field potential, and $G_5$ corresponds to the five-dimensional Newton constant. The explicit expressions of $f(\phi)$ and $V(\phi)$ are determined by solving the equations of motion (EoMs). To investigate the thermodynamic properties of quantum chromodynamics (QCD) in greater detail, we convert the action from the string frame to the Einstein frame via the following transformation relations,
\begin{equation}
\begin{split}
\phi_s=\sqrt{\frac{3}{8}} \phi, \quad &g_{\mu \nu}^s=g_{\mu \nu} e^{\sqrt{\frac{2}{3}} \phi}, \quad f_s\left(\phi_s\right)=f(\phi) e^{\sqrt{\frac{2}{3}} \phi},\\
&V_s\left(\phi_s\right)=e^{-\sqrt{\frac{2}{3}} \phi} V(\phi).
\end{split}
\end{equation}
Upon transitioning to the Einstein frame, the action assumes the form presented below.
\begin{equation}
\begin{split}
S_b & =\frac{1}{16 \pi G_5} \int d^5 x \sqrt{-g}\\
&\quad\times\left[R-\frac{f(\phi)}{4} F^2-\frac{1}{2} \partial_\mu \phi \partial^\mu \phi-V(\phi)\right].
\end{split}
\end{equation}
The equations of motion derived from the above action take the following form:
\begin{equation}
\begin{gathered}
R_{M N}-\frac{1}{2} g_{M N} R-T_{M N}=0, \\
\nabla_M\left[f(\phi) F^{M N}\right]=0, \\
\partial_M\left[\sqrt{-g} \partial^M \phi\right]-\sqrt{-g}\left(\frac{\partial V}{\partial \phi}+\frac{F^2}{4} \frac{\partial f}{\partial \phi}\right)=0,
\end{gathered}
\end{equation}
where
\begin{equation}
\begin{aligned}
T_{M N}&=\frac{1}{2}\left(\partial_M \phi \partial_N \phi-\frac{1}{2} g_{M N}(\partial \phi)^2-g_{M N} V(\phi)\right)\\
&\quad+\frac{f(\phi)}{2}\left(F_{M P} F_N^P-\frac{1}{4} g_{M N} F^2\right) .
\end{aligned}
\end{equation}
The metric is assumed to take the following form:
\begin{equation}
d s^2=\frac{L^2 e^{2 A(z)}}{z^2}\left[-g(z) d t^2+\frac{d z^2}{g(z)}+d \vec{x}^2\right],
\end{equation}
In this setup, the five-dimensional holographic coordinate is denoted by $z$, and the radius $L$ of the $\mathrm{AdS}_5$ spacetime is set to unity. Applying the above metric ansatz, the equations of motion and the associated constraints for the background fields are obtained, as presented below.
\begin{equation}\label{7}
\begin{footnotesize}
\begin{gathered}
\phi^{\prime \prime}+\phi^{\prime}\left(-\frac{3}{z}+\frac{g^{\prime}}{g}+3 A^{\prime}\right)-\frac{L^2 e^{2 A}}{z^2 g} \frac{\partial V}{\partial \phi}+\frac{z^2 e^{-2 A} A_t^{\prime 2}}{2 L^2 g} \frac{\partial f}{\partial \phi}=0,
\end{gathered}
\end{footnotesize}
\end{equation}
\begin{equation}\label{8}
A_t^{\prime \prime}+A_t^{\prime}\left(-\frac{1}{z}+\frac{f^{\prime}}{f}+A^{\prime}\right)=0,
\end{equation}
\begin{equation}\label{9}
g^{\prime \prime}+g^{\prime}\left(-\frac{3}{z}+3 A^{\prime}\right)-\frac{e^{-2 A} A_t^{\prime 2} z^2 f}{L^2}=0,
\end{equation}
\begin{equation}\label{10}
\begin{split}
&A^{\prime \prime}+\frac{g^{\prime \prime}}{6 g}+A^{\prime}\left(-\frac{6}{z}+\frac{3 g^{\prime}}{2 g}\right)\\
&-\frac{1}{z}\left(-\frac{4}{z}+\frac{3 g^{\prime}}{2 g}\right)+3 A^{\prime 2} +\frac{L^2 e^{2 A} V}{3 z^2 g}=0,
\end{split}
\end{equation}
\begin{equation}\label{11}
A^{\prime \prime}-A^{\prime}\left(-\frac{2}{z}+A^{\prime}\right)+\frac{\phi^{\prime 2}}{6}=0,
\end{equation}
Only four of these five equations are linearly independent. The horizon, located at $z=z_h$, provides the boundary conditions for the EoMs (Eq.~\ref{7}--\ref{11}). As the black hole boundary, this horizon physically introduces the temperature and is expressed as follows:
\begin{equation}
A_t\left(z_h\right)=g\left(z_h\right)=0.
\end{equation}
As the infrared (IR) boundary is approached at $z \rightarrow z_{h}$, the string-frame metric is constrained to converge to $\rm AdS_5$. The conditions at the ultraviolet (UV) cutoff, located at $z=0$, are given by:
\begin{equation}
\begin{small}
A(0)=-\sqrt{\frac{1}{6}} \phi(0) =0,  g(0)=1,  A_t(0)=\mu+\rho^{\prime} z^2+\cdots.
\end{small}
\end{equation}
In this work, $\mu$ denotes the baryon chemical potential, and $\rho^{\prime}$ is a quantity proportional to the net baryon number density. The relation between $\mu$ and the quark chemical potential is given by $\mu = 3\mu_q$. The detailed procedure for evaluating the baryon number density can be found in Refs.~\cite{Critelli:2017oub,Zhang:2022uin}.
\begin{equation}
\begin{aligned}
\rho & =\left|\lim _{z \rightarrow 0} \frac{\partial \mathcal{L}}{\partial\left(\partial_z A_t\right)}\right| \\
& =-\frac{1}{16\pi G_5} \lim _{z \rightarrow 0}\left[\frac{\mathrm{e}^{A(z)}}{z} f(\phi) \frac{\mathrm{d}}{\mathrm{d} z} A_t(z)\right].
\end{aligned}
\end{equation}
Within the Einstein frame, the EoMs can be explicitly derived from the Lagrangian density $\mathcal{L}$ by means of analytical methods.
\begin{align}
\begin{split}
\phi^{\prime}(z) & =\sqrt{-6\left(A^{\prime \prime}-A^{\prime 2}+\frac{2}{z} A^{\prime}\right)}, \\
A_t(z) & =\sqrt{\frac{-1}{\int_0^{z_h} y^3 e^{-3 A} d y \int_{y_g}^y \frac{x}{e^A f} d x}} \int_{z_h}^z \frac{y}{e^A f} d y, \\
g(z) & =1-\frac{\int_0^z y^3 e^{-3 A} d y \int_{y_g}^y \frac{x}{e^A f} d x}{\int_0^{z_h} y^3 e^{-3 A} d y \int_{y_g}^y \frac{x}{e^A f} d x}, \\
V(z) & =-3 z^2 g e^{-2 A}\bigg[A^{\prime \prime}+3 A^{\prime 2}+\left(\frac{3 g^{\prime}}{2 g}-\frac{6}{z}\right) A^{\prime}\\
& \quad -\frac{1}{z}\left(\frac{3 g^{\prime}}{2 g}-\frac{4}{z}\right)+\frac{g^{\prime \prime}}{6 g}\bigg].
\end{split}
\end{align}
The sole undetermined constant $y_g$ could be related to the chemical potential $\mu$ in a particular way. This relation is established by performing an expansion of the field $A_t(z)$ in the neighborhood of the boundary $z=0$, leading to
\begin{align}
\begin{split}
A_t(0) &=\sqrt{\frac{-1}{\int_0^{z_h} y^3 e^{-3 A} d y \int_{y_g}^y \frac{x}{e^A f} d x}}\\
&\quad \times\bigg(-\int_0^{z_h} \frac{y}{e^A f} d y+\frac{1}{e^{A(0)} f(0)} z^2+\cdots\bigg).
\end{split}
\end{align}
According to the AdS/CFT correspondence, the chemical potential of the system is given by:
\begin{small}
\begin{equation}\label{2-20}
\mu=-\sqrt{\frac{-1}{\int_0^{z_h} y^3 e^{-3 A(y)}(\int_{y_g}^y \frac{x}{e^{A(x)}f(x)}dx)dy}} \int_0^{z_h} \frac{y}{e^{A(y)}f(y)}dy.
\end{equation}
\end{small}
We employ a general ansatz for the gauge kinetic function $f(z)$:
\begin{equation}\label{fff}
f(z)=\frac{ke^{cz^{2}+hz^{4}+nz^{6}-A(z)}}{-2c-4hz^{2}-6nz^{4}}.
\end{equation}
With this setup, Eq.~(\ref{2-20}) reduces to a simpler form, and the dependence of the integration constant $y_g$ on the chemical potential $\mu$ is expressed as follows:
\begin{equation}
e^{-c y_g^2-hy_{g}^{4}-ny_{g}^{6}}=\frac{I_{2}(0,z_{h})}{I_{1}(0,z_{h})}+\frac{(e^{-cz_{h}^{2}-hz_{h}^{4}-nz_{h}^{6}}-1)^{2}}{k\mu^2I_{1}(0,z_{h})}.
\end{equation}
where
\begin{equation}
\begin{split}
I_{1}(a,b)&=\int_{a}^{b}y^{3}e^{-3A(y)}dy\\
I_{2}(a,b)&=\int_{a}^{b}y^{3}e^{-3A(y)-cy^{2}-hy^{4}-ny^{6}}dy\\
\end{split}
\end{equation}
Hence, the solutions are simplified to the forms given below
\begin{equation}
\begin{aligned}
g(z)&=1-\frac{1}{I_{1}(0,z_{h})}\Bigg[I_{1}(0,z)-\frac{\mu^2k}{\Delta^2} \operatorname{det} \mathcal{G}\Bigg],\\
\phi^{\prime}(z) & =\sqrt{6\left(A^{\prime 2}-A^{\prime \prime}-2 A^{\prime} / z\right)}, \\
A_t(z) & =\mu \frac{e^{-c z^2-hz^{4}-nz^{6}}-e^{-c z_h^2-hz_{h}^{4}-nz_{h}^{6}}}{1-e^{-c z_h^2-hz_{h}^{4}-nz_{h}^{6}}}, \\
V(z) & =-\frac{3 z^2 g e^{-2 A}}{L^2}\bigg[A^{\prime \prime}+A^{\prime}\left(3 A^{\prime}-\frac{6}{z}+\frac{3 g^{\prime}}{2 g}\right)\\
&\quad-\frac{1}{z}\left(-\frac{4}{z}+\frac{3 g^{\prime}}{2 g}\right)+\frac{g^{\prime \prime}}{6 g}\bigg],
\end{aligned}
\end{equation}
with
\begin{equation}
\Delta=e^{-cz_{h}^{2}-hz_{h}^{4}-nz_{h}^{6}}-1,
\end{equation}
and
\begin{equation}
\operatorname{det} \mathcal{G}=\left|\begin{array}{ll}
I_{1}(0,z_{h}) & I_{2}(0,z_{h}) \\
I_{1}(z_{h},z) & I_{2}(z_{h},z)
\end{array}\right|.
\end{equation}
For this black hole solution, the Hawking temperature~\cite{Natsuume:2014sfa} and the associated entropy are expressed as follows:
\begin{small}
\begin{align}
\begin{split}
T &=\frac{\left|g^{\prime}(z)\right|}{4\pi}\\
& =\frac{z_h^3 e^{-3 A\left(z_h\right)}}{4\pi I_{1}(0,z_{h})}\bigg[1-\frac{\mu^2 k}{\Delta^{2}}\Big((\Delta+1)I_{1}(0,z_{h})-I_{2}(0,z_{h})\Big)\bigg],
\end{split}
\end{align}
\end{small}
and
\begin{equation}\label{eq2-28}
S=\frac{e^{3 A\left(z_h\right)}}{4 G_5 z_h^3}.
\end{equation}
The system is solved analytically by employing the following ansatz:
\begin{equation}\label{AAA}
A(z)= d*\ln(a z^2 + 1) + d*\ln(b z^4 + 1).
\end{equation}
The role of $A(z)$ is to replicate the appropriate entropy dependence and to enforce restrictions on the temperature-dependent model motivated by Refs.~\cite{Dudal:2017max,Chen:2020ath,Chen:2024ckb,Chen:2024mmd}. In the string frame, $A_s(z)$ takes the following form:
\begin{equation}
A_s(z)=A(z)+\sqrt{\frac{1}{6}} \phi(z).
\end{equation}
The first four parameters ($a$, $b$, $d$, $G_5$) are fixed by a bayesian analysis technique that matches the holographic equation of state to LQCD results, whereas the remaining parameters ($c$, $k$, $h$, and $n$) are constrained using the baryon number susceptibility and baryon number density with bayesian analysis, as detailed in the next section. Once both the entropy and the density $\rho$ have been obtained, the free energy and pressure can be computed by integrating the thermodynamic relation:
\begin{align}
dF=-dP=-SdT-\rho d\mu
\end{align}
and
\begin{equation}
\begin{aligned}
F&=-\int S d T-\int \rho d \mu\\
&=-\int S\left(\frac{\partial T}{\partial z_h} d z_h+\frac{\partial T}{\partial \mu} d \mu\right)-\int\rho d \mu\\
&= \int_{z_h}^\infty S\frac{\partial T}{\partial z_h} d z_h-\int_0^\mu (\frac{\partial T}{\partial \mu} d \mu\ +\rho) d \mu.
\end{aligned}
\end{equation}
The free energy is normalized to vanish in the limit $z_h \rightarrow \infty$. Consequently, the energy density of the system can be expressed as:
\begin{equation}
\epsilon=-P+S T+\mu \rho.
\end{equation}
At finite chemical potential, the squared speed of sound can be obtained following the procedures described in \cite{Li:2020hau,Yang:2017oer,Gursoy:2017wzz}.
\begin{equation}\label{eq2-33}
C_s^2=\frac{S}{T\left(\frac{\partial S}{\partial T}\right)_\mu+\mu\left(\frac{\partial \rho}{\partial T}\right)_\mu}.
\end{equation}
Specific heat capacity refers to the amount of heat absorbed or released by a unit mass of a substance during temperature changes, which can be defined as \cite{DeWolfe:2010he}:
\begin{equation}
C_V=T \frac{\partial S}{\partial T}.
\end{equation}
Here, we introduce the normalized second-order baryon number susceptibility, which takes the form:
\begin{equation}\label{eq2-35}
\chi_{2}^{B}=\frac{1}{T^2} \frac{\partial \rho}{\partial \mu}.
\end{equation}

\section{Bayesian inference of EMD model parameters}\label{sec3}

Bayesian inference has been remarkably effective for constraining parameters in models of relativistic heavy-ion collisions \cite{Pratt:2015zsa,Novak:2013bqa,Sangaline:2015isa}. Key applications encompass the extraction of the shear viscosity to entropy density ratio ($\eta/s$) of the QGP medium \cite{Bernhard:2015hxa,Bernhard:2016tnd}, the jet transport coefficient $\hat{q}$ \cite{Xie:2022ght,Xie:2022fak,JETSCAPE:2021ehl}, the QCD equation of state \cite{OmanaKuttan:2022aml}, heavy quarkonium dissociation \cite{Zhu:2025xiz}, the transport properties of QGP \cite{Chen:2025fpd}, nucleon distributions within the nucleus \cite{Cheng:2023ucp}, collectivity arising from transverse momentum conservation (TMC) effects in small collision systems~\cite{Guo:2026zhq}, the decay of heavy and superheavy nuclei \cite{Zhu:2025ujz}, QCD phase transitions inside neutron stars \cite{Li:2025obt}, heavy quark energy loss~\cite{Guo:2026ndq}, and investigations of jet energy loss \cite{Zhu:2025edv, Xing:2023ciw, He:2018gks, Wu:2023azi}. In the present study, we employ Bayesian inference techniques to systematically examine how LQCD data \cite{HotQCD:2014kol,Bazavov:2017dus} restrict the parameter space of the EMD model. In particular, we infer the posterior distributions for the EMD model parameters $\boldsymbol{\theta}=(a,b,c,d,k,n,h,G_{5})$ and explore their impact on the predictions for the entropy, the second-order baryon susceptibility, and the squared speed of sound.

Bayesian inference provides a natural framework for parameter estimation and uncertainty quantification, with the final outcome being a posterior probability distribution that encapsulates the most probable values of the parameters. In the general formulation, the model parameters are collected into a vector $\boldsymbol{\theta}$, and the experimental measurements are represented by the vector $\mathbf{data}$. For our specific analysis, this $\mathbf{data}$ vector is taken to be the LQCD data described above. We first specify a prior distribution $P(\boldsymbol{\theta})$ encoding the initial knowledge about the parameters, and a likelihood function $P(\mathbf{data}|\boldsymbol{\theta})$ that, given the parameters $\boldsymbol{\theta}$, quantifies the agreement between the model predictions and the adopted LQCD data while accounting for all relevant uncertainties. Bayes' theorem is then applied to obtain the posterior distribution
\begin{align}
P(\boldsymbol{\theta} | \mathbf{data}) \propto P(\mathbf{data} |\boldsymbol{\theta})\,P(\boldsymbol{\theta}).
\end{align}

In this Section, we provide a comprehensive account of the bayesian analysis workflow for estimating the parameters of the EMD model. Subsection \textcolor{blue}{A} covers the model input specifications, including the selection of input parameters and their assigned distributions over the parameter space. Subsection \textcolor{blue}{B} addresses the model outputs, specifically the post‑processing of simulation results required to build a Gaussian process emulator as a surrogate model. Subsection \textcolor{blue}{C} presents the theoretical foundation of Gaussian process emulation and the practical implementation details for constructing the emulator. Subsection \textcolor{blue}{D}  then elaborates on the bayesian inference process for obtaining the model parameter, covering markov chain monte carlo (MCMC) \cite{Foreman-Mackey:2012any, Goodman:2010dyf} sampling, the construction of the posterior distribution, and uncertainty quantification.

\subsection{Parameter design of the prior distribution}\label{sec3-1}

The parameter setup for the EMD model, denoted by $\boldsymbol{\theta}=(a,b,c,d,k,n,h,G_{5})$, is outlined here. Two key steps are involved: (i) fixing the count of design points and their respective intervals; (ii) placing these points strategically across the parameter domain.

The first task concerns the determination of the number of design points---that is, the total count of parameter vectors $\boldsymbol{\theta}=(a,b,c,d,k,n,h,G_{5})$ to be generated for the training dataset required by the subsequent emulator construction. This decision must strike a delicate balance between computational feasibility and statistical reliability: the goal is to keep the overall computational cost as low as possible while simultaneously guaranteeing that the parameter space is adequately sampled and that the resulting Bayesian posterior estimates retain sufficient precision. To this end, we employ $N$ design points, arranged into a $N\times 8$ matrix $\mathbf{\Theta}=(\boldsymbol{\theta}_1,\boldsymbol{\theta}_2,\ldots,\boldsymbol{\theta}_N)$, where each row corresponds to a distinct parameter configuration. As for the permissible intervals of the individual parameters, they are chosen to be sufficiently wide in physical terms to cover all realistic scenarios, yet we deliberately refrain from imposing any unnecessary restrictions that could introduce systematic biases into the inference.

Secondly, we resort to Latin hypercube sampling (LHS) \cite{c42ec141-810d-3108-8740-320eb5d0f4b6, MORRIS1995381} to spread the design points evenly over the eight-dimensional parameter space. As a space-filling design strategy, LHS is particularly effective for drawing representative samples from high-dimensional domains. In contrast to naive random sampling, this technique divides the range of every parameter into $N$ equal segments and extracts exactly one sample from each segment, which ensures that every dimension is marginally covered in a uniform fashion. The main virtue of such a stratified scheme lies in its ability to produce a well-dispersed set of points, thereby circumventing the issues of point clustering and regional omission. Even when the available computational budget is modest and only a limited number of sample points can be afforded, LHS still manages to cover the entire parameter space efficiently. This feature renders it highly attractive for the subsequent construction of surrogate models (e.g., Gaussian process emulators), as a uniformly populated training set facilitates more faithful interpolation over the high-dimensional input space, ultimately leading to enhanced accuracy in the Bayesian inference.

For the 8-dimensional parameter space of the EMD model, $\boldsymbol{\theta} = (a, b, c, d, k, n, h, G_5)$, LHS was employed to generate 500 design points, i.e., 500 parameter sets $\boldsymbol{\theta}$, which were then organized into a $500 \times 8$ design matrix $\boldsymbol{\Theta} = (\boldsymbol{\theta}_1, \boldsymbol{\theta}_2, \dots, \boldsymbol{\theta}_{500})$. Although previous studies \cite{article} have indicated that approximately 10 design points are typically sufficient to yield acceptable computational accuracy, 500 points were adopted in this work to further improve the emulator performance, thus achieving a better trade-off between computational feasibility and statistical fidelity. Each parameter set $\theta_i$ was inserted into the EMD model to compute the corresponding observables $S/T^{3}$, $\chi_2^B$, and $C_s^2$. After post-processing (see the next subsection for details), the resulting 500 sets of outputs were compared with LQCD constraints.

\subsection{Post-processing of model outputs}\label{sec3-2}

\begin{figure*}[tbp!]
     \centering
     \includegraphics[width=1.0\textwidth]{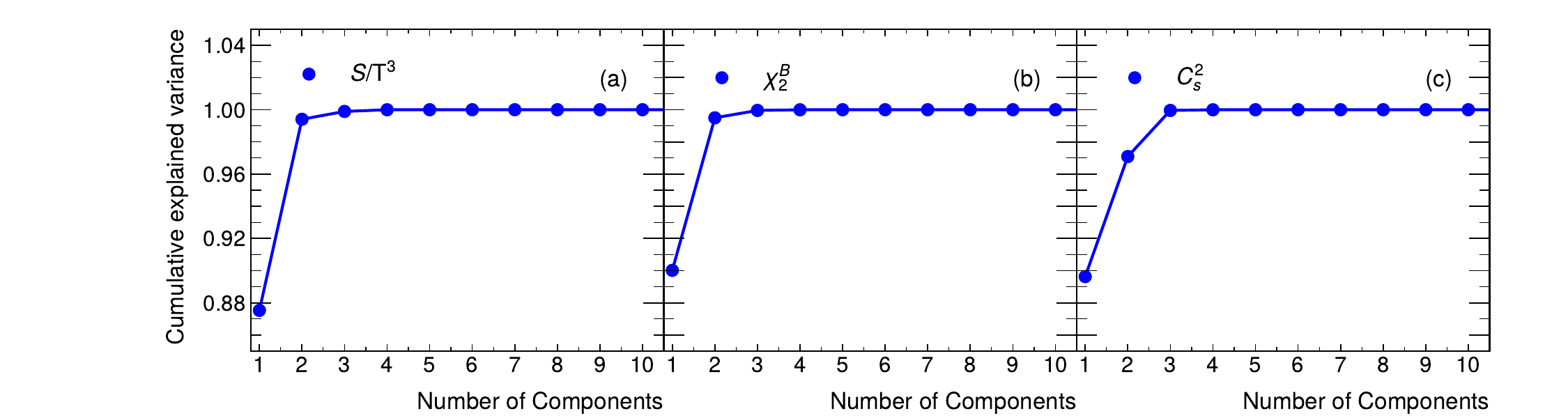}
     \caption{The cumulative explained variance as a function of the number of principal components (PCs) is shown for (a) $S/T^{3}$, (b) $\chi_{2}^{B}$, and (c) $C_{s}^{2}$. In each case, the leading three PCs account for $99\%$ of the total variance, while six PCs are kept to ensure a conservative emulation.}
     \label{fig1}
\end{figure*}

The sampled parameter sets $\mathbf{\Theta}=(\boldsymbol{\theta}_{1},\boldsymbol{\theta}_{2},\dots,\boldsymbol{\theta}_{500})$ are fed into the EMD model to compute three thermodynamic observables, namely $S/T^{3}$, $\chi_2^B$, and $C_s^2$. For each observable, the model produces 500 output vectors $\mathbf{y}=(y_1,y_2,\dots,y_m)$, where $m$ denotes the number of temperature points associated with that observable—55 for both $S/T^{3}$ and $C_s^2$, and 35 for $\chi_2^B$ (these quantities are defined in Eqs.~(\ref{eq2-28}), (\ref{eq2-33}), and (\ref{eq2-35}), respectively). Based on these outputs together with the parameter sets, independent Gaussian process emulators are constructed for each observable (see next subsection). Nevertheless, directly employing the high-dimensional output vectors for emulator construction leads to poor scalability when $m$ is large, and also overlooks the correlations among different output components. To address these drawbacks, we post-process the outputs by means of principal component analysis (PCA)~\cite{10.1162/089976699300016728}, which reduces the dimensionality and decorrelates the data while retaining the original covariance structure.

PCA maps correlated high-dimensional data onto a lower-dimensional subspace defined by orthogonal principal components, which are ranked in descending order of the variance they explain. In our application, the outputs for each observable are arranged as a $500\times m$ matrix $Y$, with rows corresponding to individual parameter samples and columns to the observable values at distinct temperature points. Before applying PCA, we standardize $Y$ by centering and scaling each column to unit variance, so that all temperature points are given equal weight in the subsequent analysis.

We perform singular value decomposition (SVD) on the standardized data, yielding the orthonormal transformation matrix $V$, whose columns $v_j$ define the principal component directions. The projection of the data onto the PC space is then given by $Z = Y V$, with $Z$ being a $500 \times m$ matrix. The rows of $Z$ index the design points, while its columns correspond to the principal components, arranged in descending order of explained variance. The first principal component captures the dominant covariance structure of $Y$, and the subsequent components encode orthogonal modes of progressively diminishing importance.

For correlated physical observables, the dominant principal components generally account for the bulk of the total variance in the system. Figure~\ref{fig1} displays the cumulative variance explained as a function of the number of retained principal components for \(S/T^3\), \(\chi_2^B\), and \(C_s^2\). Although the first three components already explain 99\% of the variance for all three observables, we opt for a more conservative choice by keeping six PCs, thereby ensuring robust emulation of any subtle structures embedded in the data. Consequently, the effective dimensionality is reduced from \(m = 55\), \(35\), and \(55\) to a common value of 6, which leads to a substantial reduction in computational expense while maintaining full predictive performance.

The PCA pipeline was built on Python's scikit-learn library, whose highly optimized numerical routines efficiently handle SVD and projection onto the principal components. After dimensionality reduction, the first six retained components, \(z_1, z_2, \dots, z_6\), are mutually uncorrelated and constitute a set of decorrelated low-dimensional target variables. These components are then employed as inputs to Gaussian process emulation, facilitating efficient bayesian inference while fully preserving the correlation structure across the original observables.

\subsection{Gaussian Processes Emulator}\label{sec3-3}

To substantially reduce the computational burden while maintaining high predictive fidelity during the subsequent large-scale exploration of the Bayesian posterior, we adopt a Gaussian process (GP) emulator \cite{williams2006gaussian} as an efficient surrogate for the full EMD model calculation.
A GP is a highly expressive, non-parametric probabilistic model that defines a distribution over functions. Its defining property is that any finite collection of function values follows a multivariate Gaussian distribution. Concretely, given the eight-dimensional input parameters $\boldsymbol{\theta}\in \mathcal{R}^8$ (with the design matrix $\mathbf{\Theta}=(\boldsymbol{\theta}_{1},\boldsymbol{\theta}_{2},...,\boldsymbol{\theta}_{500})$) and the corresponding model outputs $\mathbf{Y}=(\mathbf{y}_{1},\mathbf{y}_{2},...,\mathbf{y}_{500})$, a GP is completely characterized by the following two essential components:
\begin{itemize}
    \item \textbf{Mean function} $\mu(\boldsymbol{\theta})$: To simplify the prior specification and concentrate on the underlying variation patterns, we first center the output data ($Y\leftarrow Y - \text{mean}(Y)$) and subsequently set the mean function to zero.
    \item \textbf{Covariance kernel} $\sigma(\boldsymbol{\theta}, \boldsymbol{\theta}')$: This function captures the correlation between outputs at distinct input points, serving as the key component that encodes the dependency structure, controls function smoothness, and determines generalization capability.
\end{itemize}
Through the flexible configuration of the mean function and covariance kernel, the GP furnishes a unified and powerful framework that not only effectively models complex dependencies in the data but also provides rigorous quantification of predictive uncertainty.

We employ the squared-exponential (SE) kernel, leveraging its inherent infinite differentiability to produce exceptionally smooth functions that capture long-range trends, while its hyperparameters (e.g., the length scale) offer straightforward interpretability, thereby facilitating prior specification and physical insight into the emulation results.
\begin{align}
\sigma(\boldsymbol{\theta}, \boldsymbol{\theta}') = \exp\left(-\frac{|\boldsymbol{\theta} - \boldsymbol{\theta}'|^2}{2\ell^2}\right),
\label{eq:SE_kernel }
\end{align}
Here, the positive hyperparameter $l$ denotes the characteristic length scale that dictates the rate at which correlations between function values diminish across the input space. When the Euclidean distance between two input points is much smaller than $l$ (i.e., $||\boldsymbol{\theta}-\boldsymbol{\theta}'||\ll l$), their corresponding outputs are strongly correlated, exhibiting nearly identical local behavior. In contrast, when the distance substantially exceeds $l$ ($||\boldsymbol{\theta} - \boldsymbol{\theta}'||\gg l$), the correlation decays to near zero, rendering the outputs approximately independent. This length scale thus provides an intuitive control over the local smoothness and global variability of the inferred function.

By the definition of the Gaussian process, the collection of outputs $\mathbf{Y}$ over the complete set of $N=500$ training inputs follows a joint multivariate Gaussian distribution.
\begin{align}
\begin{split}
\mathbf{Y} \sim \mathcal{N}\left(\mathbf{0},\mathbf{K}\right),\\ 
\end{split}
\end{align}
\begin{align}
\begin{split}
\mathbf{K} =
\begin{pmatrix}
\sigma(\boldsymbol{\theta}_1, \boldsymbol{\theta}_1) & \sigma(\boldsymbol{\theta}_1, \boldsymbol{\theta}_2) & \cdots & \sigma(\boldsymbol{\theta}_1, \boldsymbol{\theta}_{500}) \\
\vdots & \vdots & \ddots & \vdots \\
\sigma(\boldsymbol{\theta}_{500}, \boldsymbol{\theta} _1) & \sigma(\boldsymbol{\theta}_{500}, \boldsymbol{\theta}_2) & \cdots & \sigma(\boldsymbol{\theta}_{500}, \boldsymbol{\theta}_{500})
\end{pmatrix},
\label{eq:GP_prior }
\end{split}
\end{align}
Here, $\mathbf{K} \in \mathbb{R}^{500\times 500}$ is the covariance matrix whose $(i,j)$-th entry is given by the pre-specified kernel function evaluated at the training input pair $(\boldsymbol{\theta}_i, \boldsymbol{\theta}_j)$, fully encoding the pairwise correlation structure across all training inputs.

After training, for a given set of test inputs $\mathbf{\Theta}_{*}$, the Gaussian process yields a posterior predictive distribution for the outputs $\mathbf{Y}_{*}$ that is also Gaussian, expressed as follows:
\begin{align}
\mathbf{Y}_{*} |\mathbf{\Theta_{*}}, \mathbf{\Theta}, \mathbf{Y} &\sim \mathcal{N}\left(\boldsymbol{\mu}_{*}, \mathbf{\Sigma}*\right),
\label{eq:GP_posterior}
\end{align}
\begin{align}
\boldsymbol{\mu}_{*} &= \mathbf{K}_{*}\mathbf{K}^{-1} \mathbf{Y},
\label{eq:GP_mean} \
\end{align}
\begin{align}
\mathbf{\Sigma}* &= \mathbf{K}_{**} -  \mathbf{K}_{*} \mathbf{K}^{-1} \mathbf{K}^\mathrm{T}_{*},
\label{eq:GP_cov}
\end{align}
Here, $\mathbf{K}_{*}$ denotes the cross-covariance between test and training inputs, i.e., $\mathbf{K}_{*} = \sigma(\mathbf{\Theta}_*, \mathbf{\Theta})$, and $\mathbf{K}_{**}$ denotes the covariance among the test inputs, $\mathbf{K}_{**} = \sigma(\mathbf{\Theta}_*, \mathbf{\Theta}_*)$.

We regard the EMD model as an underlying latent function \(f(\boldsymbol{\theta})\) that depends on eight input parameters \(\boldsymbol{\theta} = (a,b,c,d,k,n,h,G_5)\). This function maps the inputs to a set of observables, specifically the first six principal components of \(S/T^{3}\), \(\chi_2^B\), and \(C^2_s\). The training of the Gaussian process emulator is carried out in three steps: (1) center the output matrix \(\mathbf{Y}\) by subtracting the empirical mean from each observable; (2) estimate the kernel length-scale \(l\) through maximization of the marginal likelihood; and (3) apply Eqs.~(\ref{eq:GP_posterior})--(\ref{eq:GP_cov}) to compute the predictive distribution for \(\mathbf{Y}_*\) at unseen test inputs \(\mathbf{\Theta}_*\).

By delivering both rapid interpolation over the 8-dimensional parameter space and principled uncertainty quantification, this framework provides an essential foundation for Bayesian parameter inference.
It should be noted that the results predicted by the GP emulator are in the principal component space. During the Bayesian inference, we need to inversely transform the results from the principal component space back to the real physical space, as will be explained in the next subsection.

\subsection{MCMC sampling of the posterior distribution}\label{sec3-4}

We are now able to constrain the EMD model using LQCD data (also referred to as evidence in bayesian inference), that is, to obtain numerical estimates of the model parameters while simultaneously quantifying their uncertainties. This is essentially an inverse problem, where the central task is to infer the model input parameters or physical factors~\cite{Pang:2016vdc,OmanaKuttan:2020btb,Jiang:2021gsw,Aarts:2025gyp} from the available LQCD data.
Within the bayesian inference framework, the parameter estimation is derived from the posterior distribution $P(\boldsymbol{\theta} | \mathbf{data}) \propto P(\mathbf{data} |\boldsymbol{\theta})\,P(\boldsymbol{\theta})$, which merges prior information with the likelihood function built from observational data, thereby yielding more reliable parameter estimates.

In the present work, $\boldsymbol{\theta}$ denotes the model parameters, while $\mathbf{data}$ stands for the LQCD data. This relation is Bayes' theorem: the left-hand side, the posterior distribution $P(\boldsymbol{\theta} | \mathbf{data})$, characterizes the probability distribution of the parameters given the observed data; the right-hand side $\mathit{P}(\mathbf{data}|\boldsymbol{\theta})$ is the likelihood function, which quantifies how probable it is to obtain the chosen evidence (the LQCD data) at specific parameter values; and $\mathit{P}(\boldsymbol{\theta})$ represents the prior distribution, encoding our initial beliefs or assumptions about the parameters before the data are taken into account.

In what follows, we present the specific definitions of the prior distribution and the likelihood function, and discuss how to perform efficient sampling from the posterior distribution via MCMC methods.

The prior distribution $\mathit{P}(\boldsymbol{\theta})$ reflects the initial beliefs about the parameters prior to data observation. In situations lacking specific information, a uniform prior is a common choice, which simply assigns a constant probability to $\mathit{P}(\boldsymbol{\theta})$. In the present study, we take each parameter to be uniformly distributed over a bounded design interval; consequently, the Gaussian process emulator only yields reliable predictions within this region. Hence, the prior can be set to a constant inside the design domain and to zero outside it, serving both as a practical restriction and a compact summary of prior knowledge.

\begin{align}\label{eq:3-8}
P(\mathbf{\boldsymbol{\theta}}) \propto 
\begin{cases} 
1 & \text{if } \min(\boldsymbol{\theta}_i) \leq \boldsymbol{\theta}_i \leq \max(\boldsymbol{\theta}_i) \text{ for all } i, \\
0 & \text{else.}
\end{cases}
\end{align}

Defining the prior as zero beyond the design boundary constitutes a strict and conservative choice, essentially eliminating any probability that the actual parameter values reside outside the prescribed limits. Nevertheless, this inflexibility can lead to the exclusion of legitimate parameter regions, thereby reducing the model's adaptability and forecasting strength. A more prudent alternative is to employ broader design ranges, which allow for greater uncertainty in parameter estimation. In this way, the model maintains its generality and accommodates a wider array of plausible parameter values, ultimately improving both its resilience and practical applicability.

\begin{figure}[h]
    \centering
    \includegraphics[width=0.40\textwidth]{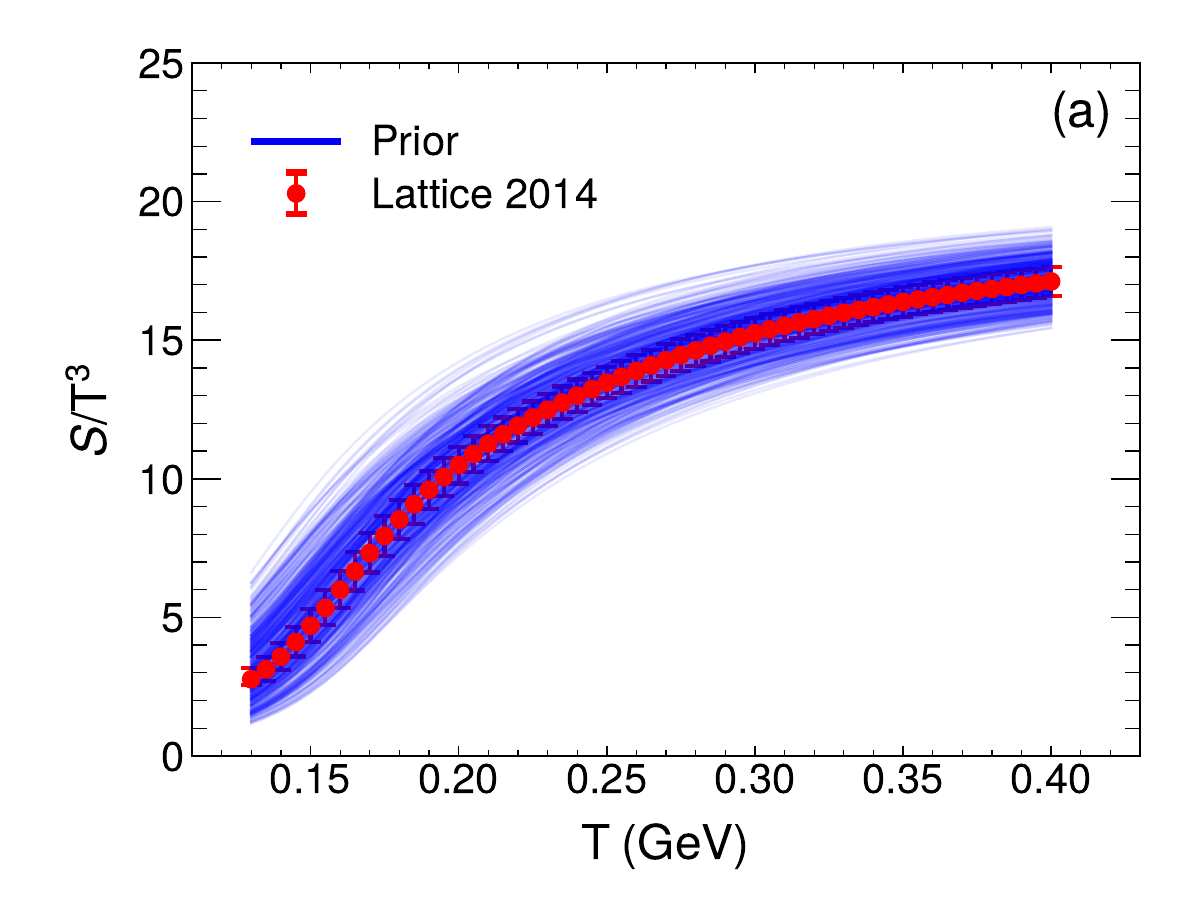}
    \includegraphics[width=0.40\textwidth]{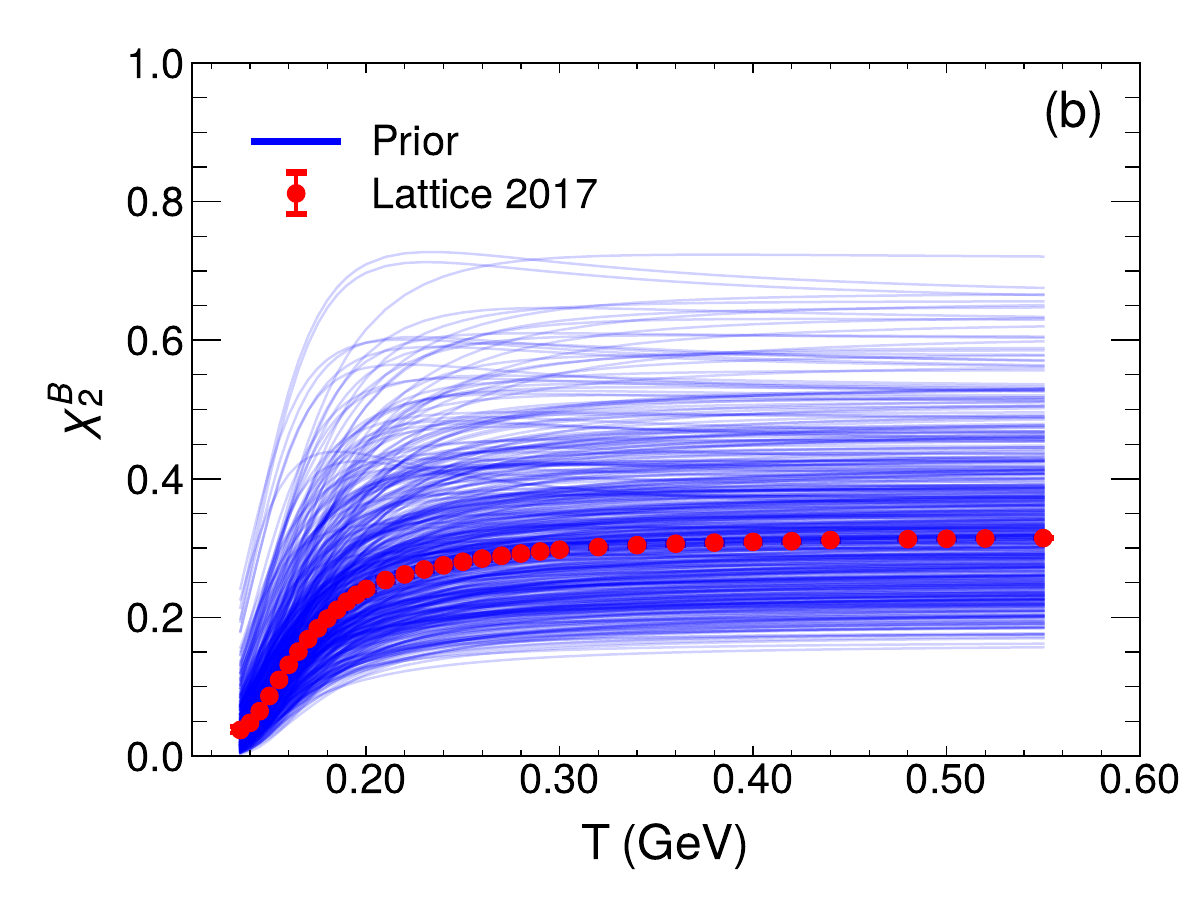}
    \includegraphics[width=0.40\textwidth]{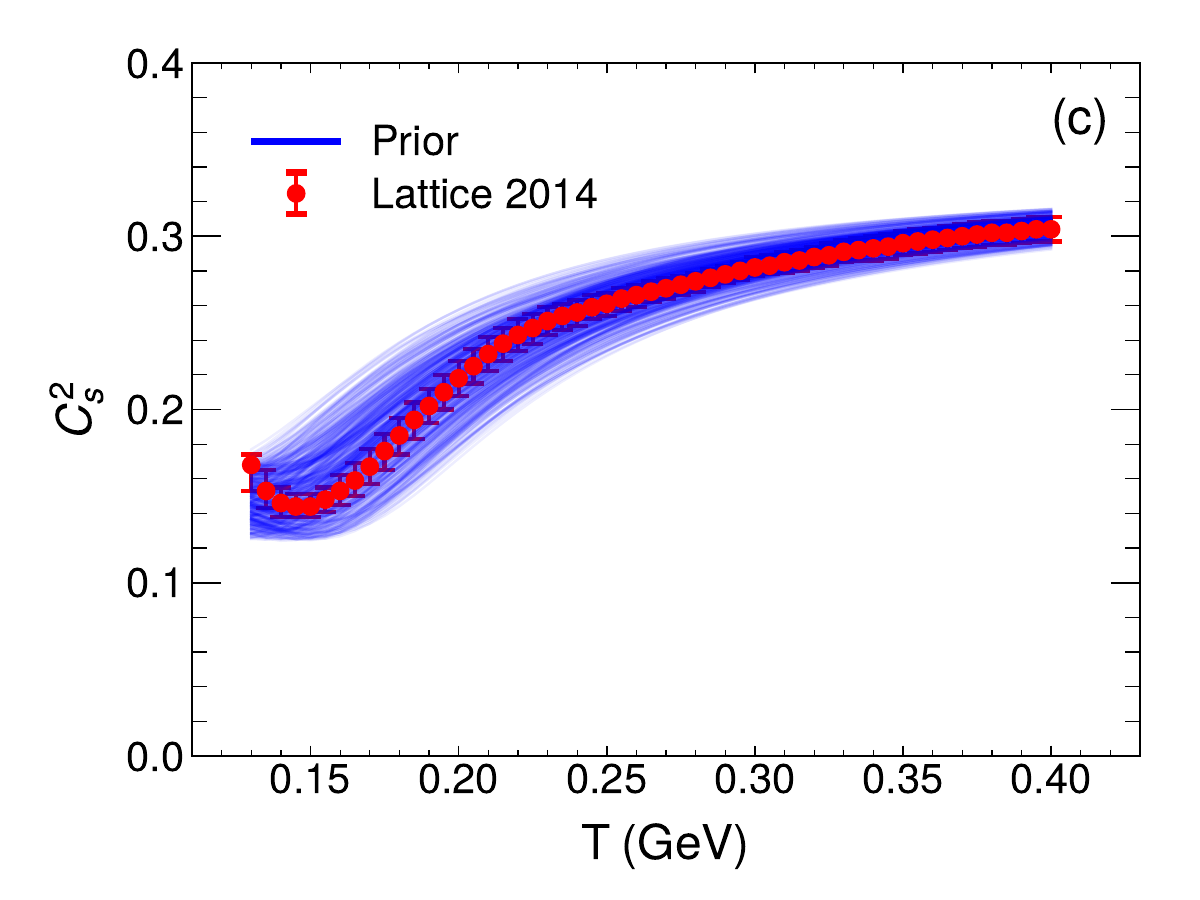}
     \caption{(a) Prior distribution of entropy scaled by temperature cubed, shown as a function of temperature. (b) Prior distribution of the baryon number susceptibility with respect to temperature. (c) Prior distribution of the speed of sound squared against temperature. The solid blue curves show the prior calculations obtained with the EMD model, whereas the red symbols with error bars correspond to the LQCD data from Ref.\cite{HotQCD:2014kol,Bazavov:2017dus}.}
     \label{fig:prior}
\end{figure}

In this study, the prior ranges for the model parameters $\boldsymbol{\theta}=(a,b,c,d,k,n,h,G_{5})$ are fully listed in Table \ref{Table:prior}. Within these intervals, we drew 500 parameter sets using the LHS method described earlier, and then fed them into the EMD model to produce theoretical predictions for $S/T^{3}$, $\chi_{2}^{B}$, and $C_{s}^{2}$, which are visually displayed in figure~\ref{fig:prior}.

\begin{table}[h]
\centering
\begin{tabular}{|c|c|c|}
  \hline
  \multicolumn{3}{|c|}{Prior} \\ \hline
  Parameter & min & max \\ \hline
  $a$ & 0.18 & 0.38 \\ \hline
  $b$ & 0.014 & 0.038 \\ \hline
  $c$ & -0.15 & -0.07 \\ \hline
  $d$ & -0.15 & -0.10 \\ \hline
  $k$ & 0.0498 & 0.1003 \\ \hline
  $n$ & -0.0025 & -0.0013 \\ \hline
  $h$ & 0.005 & 0.013 \\ \hline
  $G_{5}$ & 0.37 & 0.42 \\ \hline
\end{tabular}
\caption{The prior parameter ranges employed in the EMD model, $\text{``min''}$  denotes the minimum value of the parameter, and $\text{``max''}$  denotes the maximum value of the parameter.} 
\label{Table:prior}
\end{table}

We now turn to the likelihood function $\mathit{P}(\mathbf{data}|\boldsymbol{\theta})$, which represents the probability of actually obtaining the observed data for a given set of model parameters $\boldsymbol{\theta}$. Owing to the central limit theorem, this likelihood takes a Gaussian form, with the chi-square value entering as the exponential term:

\begin{align}\label{eq:likelihood}
\mathit{P}(\mathbf{data} |\boldsymbol{\theta})=\underset{i}{\prod}\frac{1}{\sqrt{2\pi}\sigma_{i}}e^{-\frac{[y_{i}(\boldsymbol{\theta})-y_{i}^{\text{LQCD}}]^{2}}{2\sigma_{i}^{2}}}
\end{align}

Here, $y_{i}(\boldsymbol{\theta})$ and $y_i^{\text{LQCD}}$ correspond, respectively, to the observable predictions from the GP emulator and the LQCD results. 
It should be noted that the results predicted by the GP emulator are in the principal component space, and we need to transform them back to the real physical space to obtain $y_{i}(\boldsymbol{\theta})$.
In our analysis, the error bars are self-consistently folded into the posterior estimate via the Bayesian inference framework, directly manifesting in the $\sigma^{2}$ term within the exponential of the likelihood function (Eq.~\ref{eq:likelihood}). The total error $\sigma$ combines the uncertainties from the LQCD data themselves and from the Gaussian process emulator, i.e., $\sigma^{2} = \sigma_{\mathrm{LQCD}}^{2} + \sigma_{\mathrm{GP\;emulator}}^{2}$.

In practice, we first formulate separate likelihood functions for $S/T^{3}$, $\chi_{2}^{B}$, and $C_{s}^{2}$, and then multiply them together to obtain the joint likelihood:
\begin{align}
\mathit{P}(\mathbf{data} |\boldsymbol{\theta})=\mathit{P}(\mathbf{data}_{S/T^{3}} |\boldsymbol{\theta})\mathit{P}(\mathbf{data}_{\chi_{2}^{B}} |\boldsymbol{\theta})\mathit{P}(\mathbf{data}_{C_{s}^{2}} |\boldsymbol{\theta}).
\end{align}
The product form systematically merges the constraints from various observables, thus enabling a comprehensive evaluation of the overall parameter consistency.

To sample from the posterior distribution $\mathit{P}(\mathbf{data} |\boldsymbol{\theta})$, we adopt the Metropolis-Hastings algorithm \cite{Metropolis:1953am, Hastings:1970aa}, a simple and widely used MCMC method in bayesian inference for exploring high-dimensional parameter spaces and complex distributions. The algorithm operates iteratively: from the current position $\boldsymbol{\theta}_i$, a candidate $\boldsymbol{\theta}^{\prime}$ is randomly proposed and accepted or rejected according to the posterior probabilities of the two points. If accepted, the next state becomes $\boldsymbol{\theta}_{(i+1)}=\boldsymbol{\theta}^{\prime}$; otherwise, it stays at $\boldsymbol{\theta}_{(i+1)}=\boldsymbol{\theta}_{(i)}$. Repeating this procedure yields a sequence $\{\boldsymbol{\theta}_{(1)},\boldsymbol{\theta}_{(2)},...,\boldsymbol{\theta}_{(n)}\}$ that eventually converges to the target posterior. To suppress transient correlation effects caused by the initial parameter values, one typically discards the early ``burn-in'' portion of the chain (commonly $10\%$--$20\%$ of its length). In our calculations, we take a more conservative approach to ensure the posterior is well equilibrated: we remove the first $50\%$ of the samples and use the remainder as an approximation of the target posterior.

To improve the numerical stability and computational efficiency of high-dimensional probability evaluations, we employ the log-posterior:
\begin{align}
\mathrm{log}P(\boldsymbol{\theta}| \mathbf{data})=\mathrm{log}P(\mathbf{data}|\boldsymbol{\theta})+\mathrm{log}P(\boldsymbol{\theta})+\mathrm{const}
\end{align}
Because MCMC relies solely on probability ratios, the additive constant terms that appear after taking the logarithm of a distribution have no impact on the sampling and can therefore be dropped. Concretely, the logarithmic forms of the prior Eq.~(\ref{eq:3-8}) and the uniform likelihood Eq.~(\ref{eq:likelihood}) are written, respectively, as:
\begin{align}
\mathrm{log}P(\boldsymbol{\theta}) \propto 
\begin{cases} 
0 & \text{if } \min(\boldsymbol{\theta}_i) \leq \boldsymbol{\theta}_i \leq \max(\boldsymbol{\theta}_i) \text{ for all } i, \\
-\infty & \text{else.}
\end{cases}
\end{align}
\begin{align}
\begin{split}
\mathrm{log}P(\mathbf{data}|\boldsymbol{\theta})=-\frac{1}{2}\mathbf{y}^{T}\Sigma^{-1}\mathbf{y}-\frac{1}{2}\mathrm{log}\lvert\Sigma\rvert-\frac{1}{2}\mathrm{log}2\pi,
\end{split}
\end{align}
\begin{align}
\begin{split}
\mathbf{y}=y(\boldsymbol{\theta})-y^{\mathrm{LQCD}}.
\end{split}
\end{align}

During the burn-in phase, the Markov chain moves from arbitrary initial positions towards regions of high posterior density. The necessary burn-in length is determined by the complexity of the posterior and the efficiency of the proposal distribution. In this work, we employed 180 parallel walkers, each assigned a random initial location within the parameter space. We first ran 10,000 steps as the burn-in period and discarded all these samples—equivalent to removing the first 50\% of the entire chain—to ensure adequate convergence to the target posterior distribution. Following the burn-in, another 10,000 steps were taken to collect posterior samples. This number of samples is already adequate for producing smooth histogram visualizations, but may be somewhat more than necessary for extracting summary statistics such as medians. It should be noted that the total sample size equals the number of walkers multiplied by the number of steps.

After MCMC sampling is finished, the marginal distribution of each parameter in $\boldsymbol{\theta} = (a,b,c,d,k,n,h,G_{5})$ can be obtained. A marginal distribution is derived from the posterior by singling out a specific parameter or a subset of parameters of interest, while integrating over all remaining parameters. As an illustration, in the present model, the marginal distribution of $d$ is found by integrating out the other parameters, thereby removing their influence.
\begin{align}
P(d| \mathbf{data})=\int da db dc dk dn dh dG_5P(\boldsymbol{\theta}| \mathbf{data})
\end{align}
Taking the parameters $a$ and $d$ as an example, the joint distribution of the two can be computed by the following expression:
\begin{align}
P(a,d| \mathbf{data})=\int db dc dk dn dh dG_5P(\boldsymbol{\theta}| \mathbf{data})
\end{align}

\begin{table}[h] 
\centering 
\begin{tabular}{|c|c|c|c|} 
  \hline
  \multicolumn{4}{|c|}{Posterior  95\% CL} \\ \hline
  Parameter & min & max & MAP \\ \hline
  $a$ & 0.2307 & 0.3277 & 0.2764 \\ \hline
  $b$ & 0.0225 & 0.0338 & 0.0280 \\ \hline
  $c$ & -0.1381 & -0.0972 & -0.1207 \\ \hline
  $d$ & -0.1341 & -0.1166 & -0.1234 \\ \hline
  $k$ & 0.0626 & 0.0900 & 0.0769 \\ \hline
  $n$ & -0.0023 & -0.0017 & -0.0021 \\ \hline
  $h$ & 0.0073 & 0.0120 & 0.0099 \\ \hline
  $G_{5}$ & 0.3846 & 0.4073 & 0.3969 \\ \hline
\end{tabular}
\caption{Bayesian inference posterior estimates of the EMD model parameters, together with their 95\% CL and MAP values, $\text{``min''}$  denotes the minimum value of the parameter, and $\text{``max''}$  denotes the maximum value of the parameter.} 
\label{Table:posterior}
\end{table}

\begin{figure*}[tbp!]
     \centering
     \includegraphics[width=1.0\textwidth]{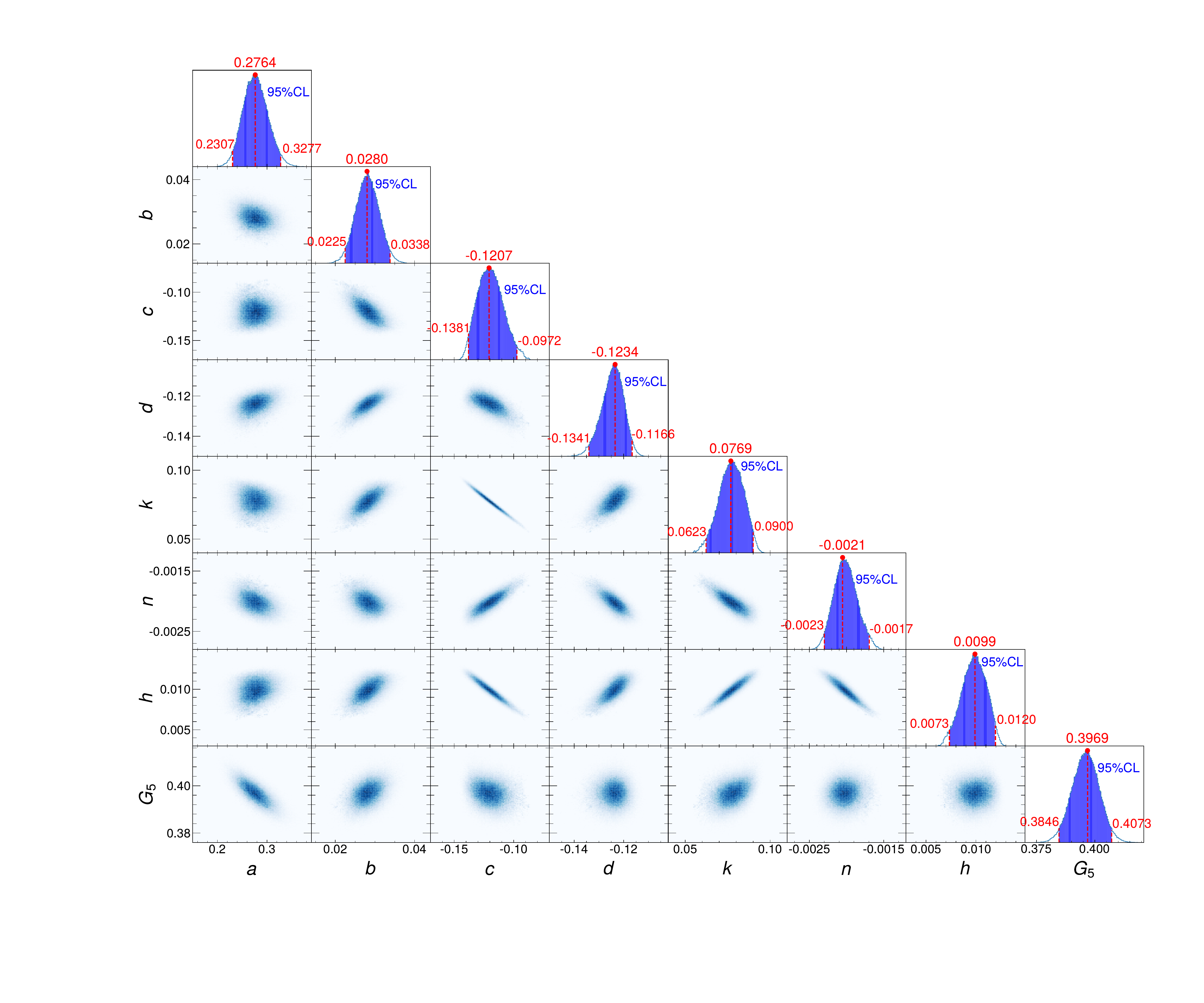}
     \caption{The diagonal panels display the posterior marginal distributions of the individual model parameters, together with the 95\% CL estimates and MAP for each parameter. The lower-triangular panels show the bivariate joint distributions for every pair of parameters.}
     \label{fig:posterior}
\end{figure*}

We take the LQCD data of $S/T^{3}$, $\chi_{2}^{B}$, and $C_{s}^{2}$ from \cite{HotQCD:2014kol,Bazavov:2017dus} as the evidence in our bayesian inference, where $C_{s}^{2}$ is specifically included to better bring out the phase transition region.
Table \ref{Table:posterior} lists the inferred parameter constraints, presenting the 95\% confidence levels (CL) and maximum a posteriori (MAP) estimates obtained from the MCMC runs.
Figure~\ref{fig:posterior} displays the posterior distributions with a corner plot (also called triangle plot), a standard visualization technique in Bayesian inference for exploring high-dimensional parameter spaces. The figure consists of diagonal and off-diagonal blocks: the diagonal panels give the marginal posterior density of each parameter after integrating out the other parameters, annotated with the 95\% CL and the MAP values; the off-diagonal panels show the bivariate joint distributions of parameter pairs through 2D density estimates, uncovering correlations or degeneracies, e.g., how different parameter combinations trade off to accommodate the observational constraints. It is clearly seen that the employed LQCD data can well calibrate every parameter in our model, yielding sharp single-peak structures, and they also indicate that certain correlations exist among all model parameters.

\section{Thermodynamics in the bayesian Holographic model}\label{sec4}

This section provides a systematic assessment of the thermodynamic predictions from our Bayesian-calibrated holographic model. In Subsection \textcolor{blue}{A}, we feed the posterior parameters inferred from the bayesian analysis into the EMD model to compute the thermodynamic quantities, and then compare these results with those from LQCD~\cite{HotQCD:2014kol,Bazavov:2017dus}, the Hadron Resonance Gas (HRG) model~\cite{Vovchenko:2019pjl}, and the Hard Thermal Loop (HTL) model~\cite{Haque:2014rua}. In Subsection \textcolor{blue}{B}, we further exploit the posterior parameters to evaluate the thermodynamic quantities at finite chemical potential and confront them with the LQCD outcomes.

\subsection{Posterior Thermodynamic Comparison to HRG and HTL}\label{sec4-1}

In this subsection, to thoroughly assess the thermodynamic performance of the bayesian holographic QCD model over a broad temperature range, we investigate its theoretical predictions at zero baryon chemical potential in both the low- and high-temperature regimes. Specifically, the various thermodynamic quantities computed from the model are carefully compared with results from LQCD, HRG, and HTL. The HRG predictions are obtained using the C++ package Thermal-FIST \cite{Vovchenko:2019pjl}, which was developed specifically for heavy-ion collision and hadronic equation-of-state studies and reliably captures the thermodynamics dominated by hadronic degrees of freedom at low temperatures. As a high-temperature reference, we also incorporate the HTL perturbative calculations \cite{Haque:2014rua}, which are based on hard thermal loop effective theory and provide perturbative estimates of thermodynamic quantities in the quark-gluon plasma phase, thereby offering an important benchmark for validating the QCD behavior of the model in the asymptotic high-temperature region.

\begin{figure*}[tbp!]
    \centering
    \begin{minipage}{0.40\textwidth}
        \includegraphics[width=\textwidth]{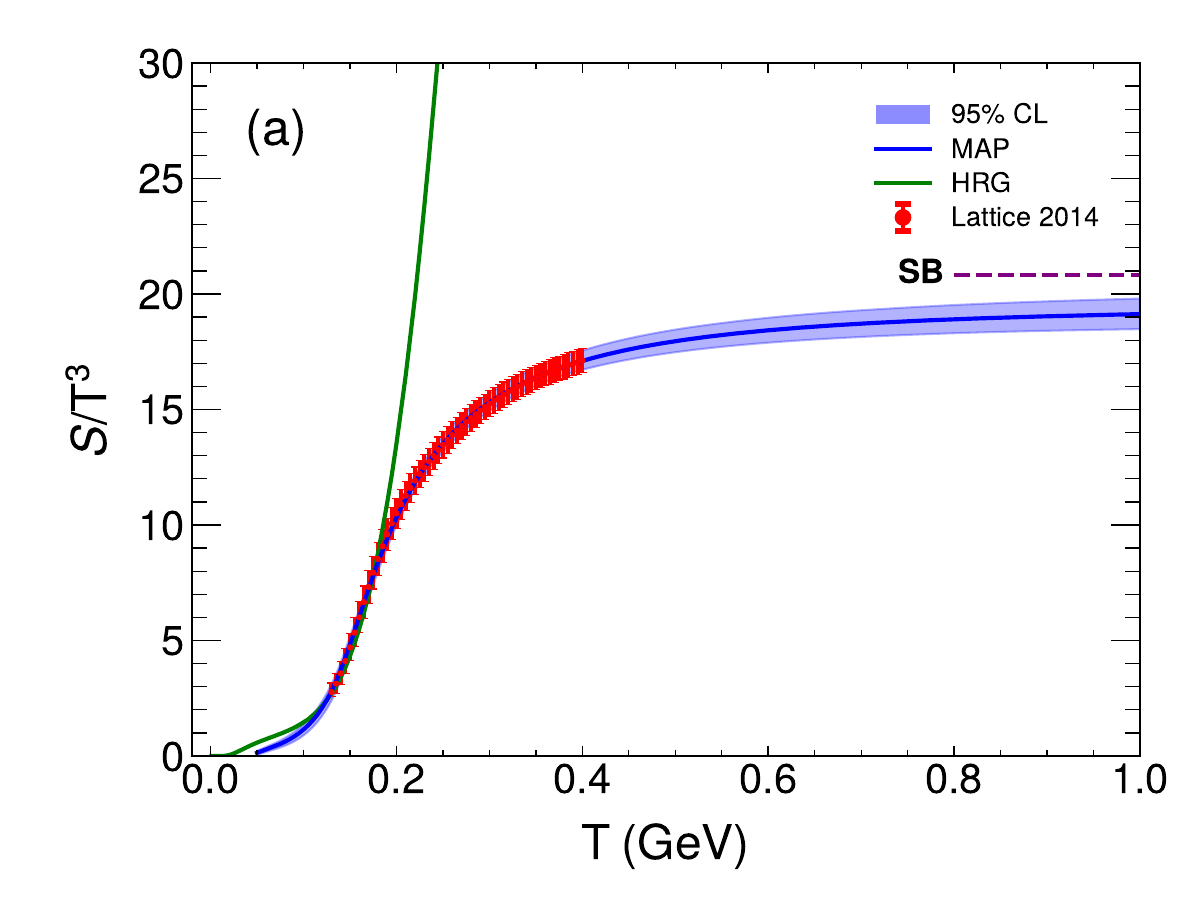}
    \end{minipage}
    \begin{minipage}{0.40\textwidth}
        \includegraphics[width=\textwidth]{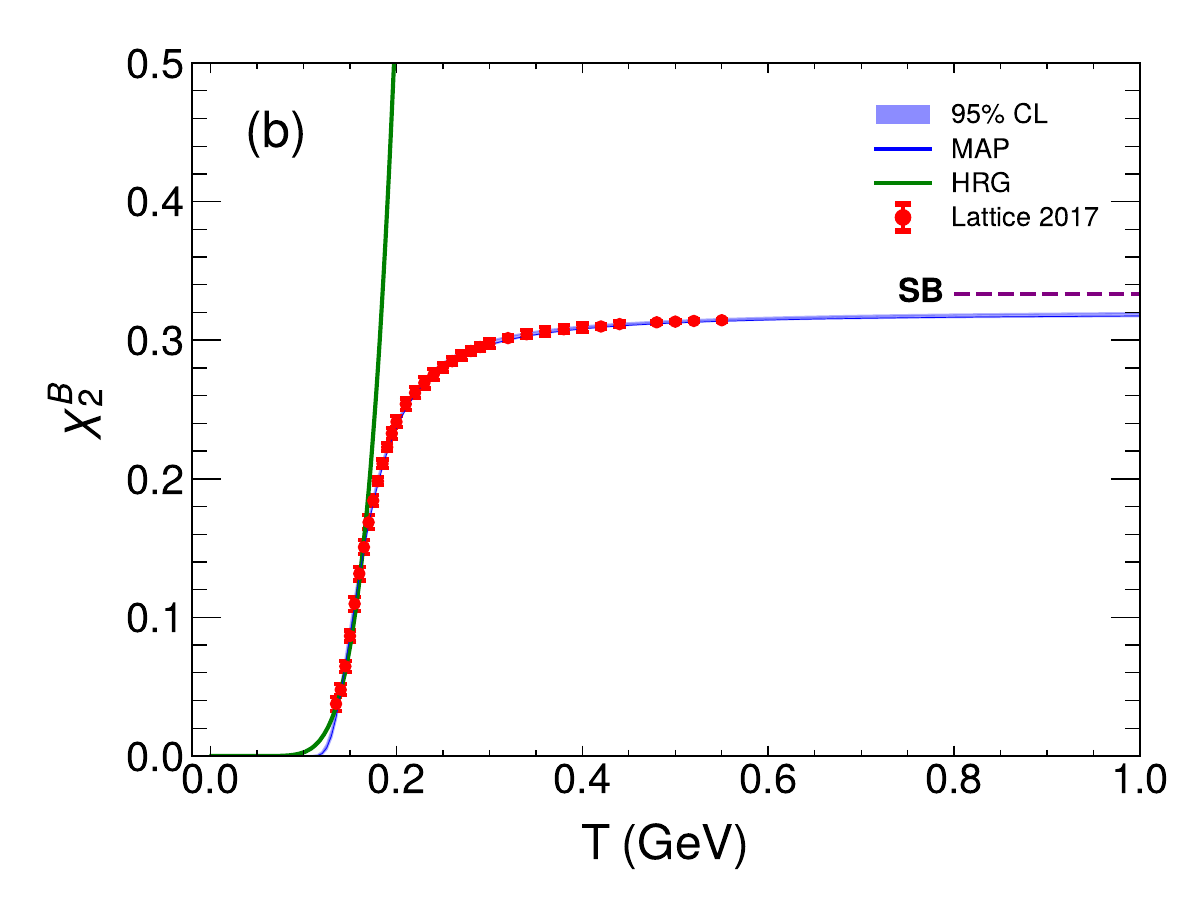}
    \end{minipage}
    \begin{minipage}{0.40\textwidth}
        \includegraphics[width=\textwidth]{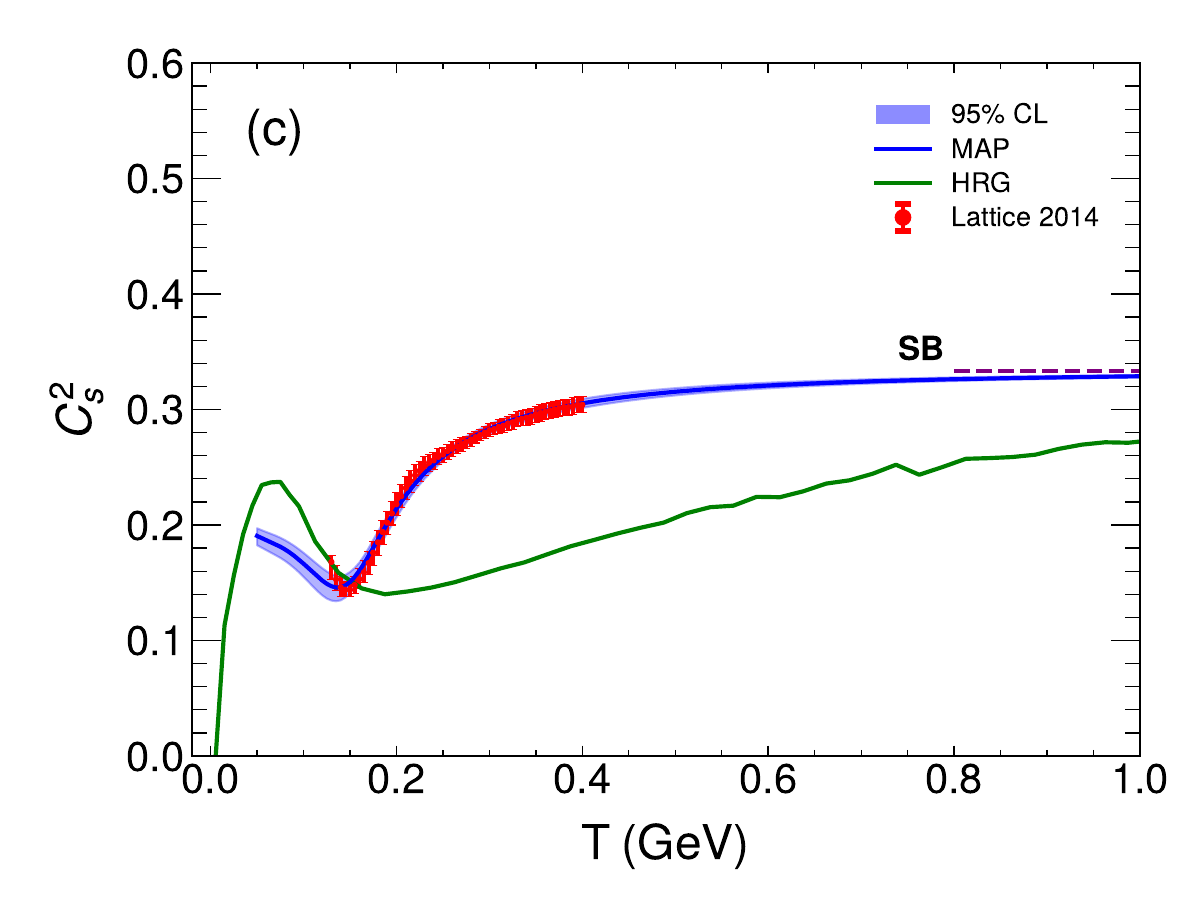}
    \end{minipage}
    \begin{minipage}{0.40\textwidth}
        \includegraphics[width=\textwidth]{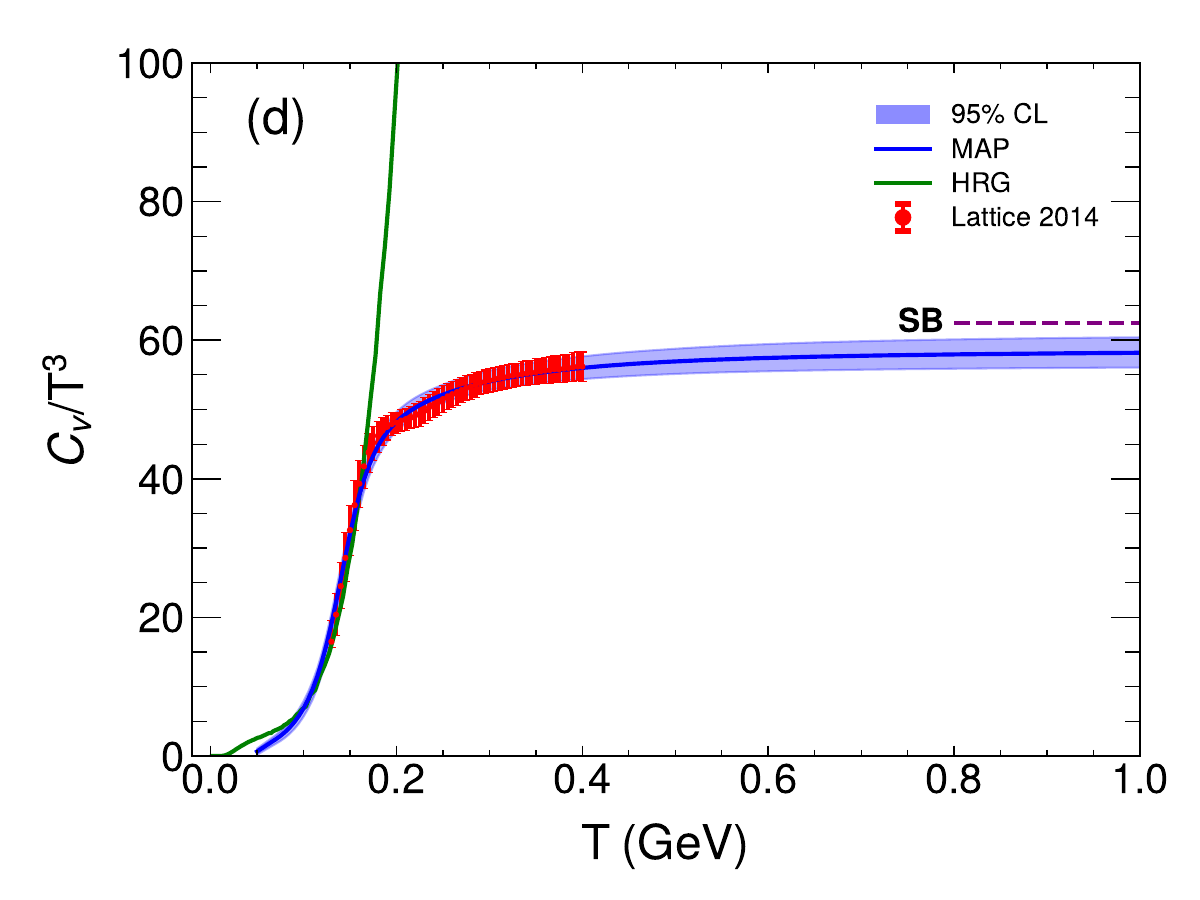}
    \end{minipage}
    \begin{minipage}{0.40\textwidth}
        \includegraphics[width=\textwidth]{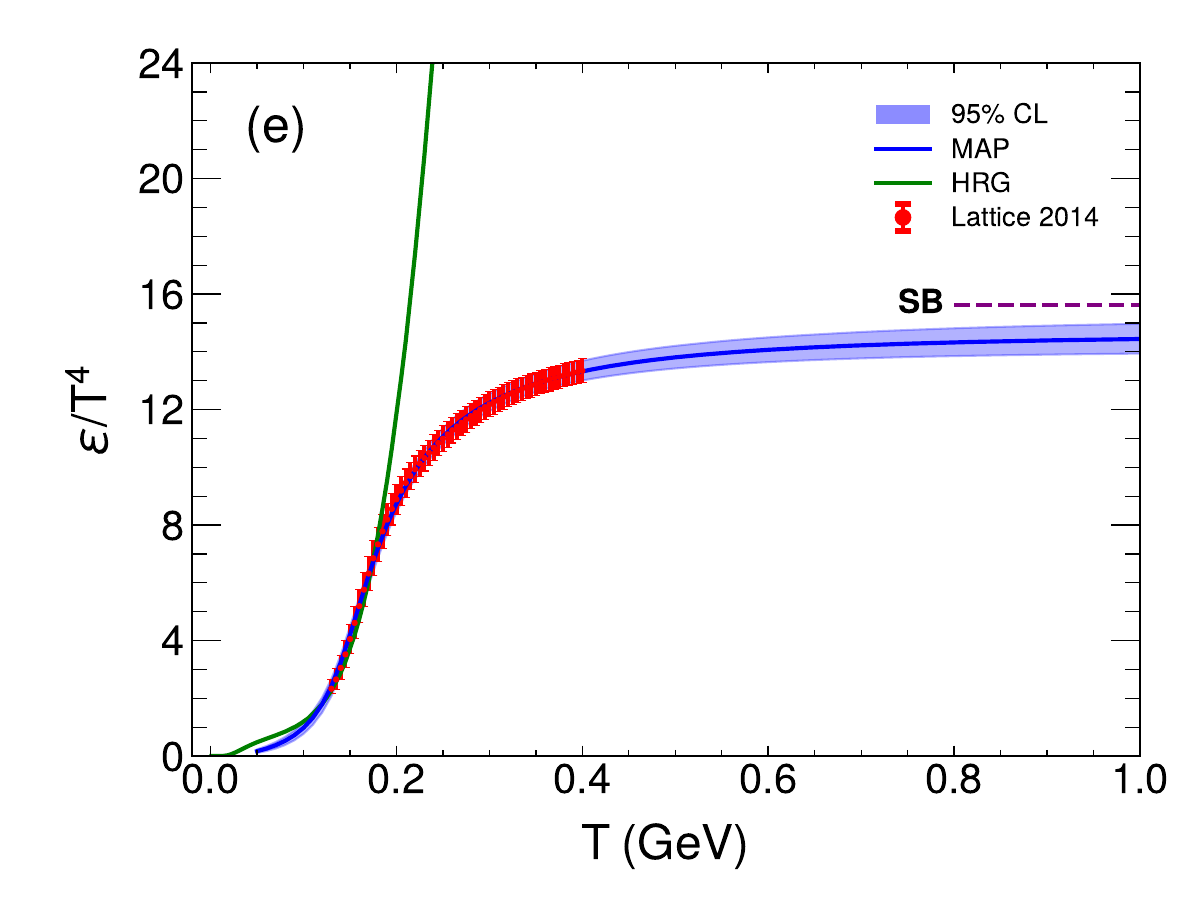}
    \end{minipage}
    \begin{minipage}{0.40\textwidth}
        \includegraphics[width=\textwidth]{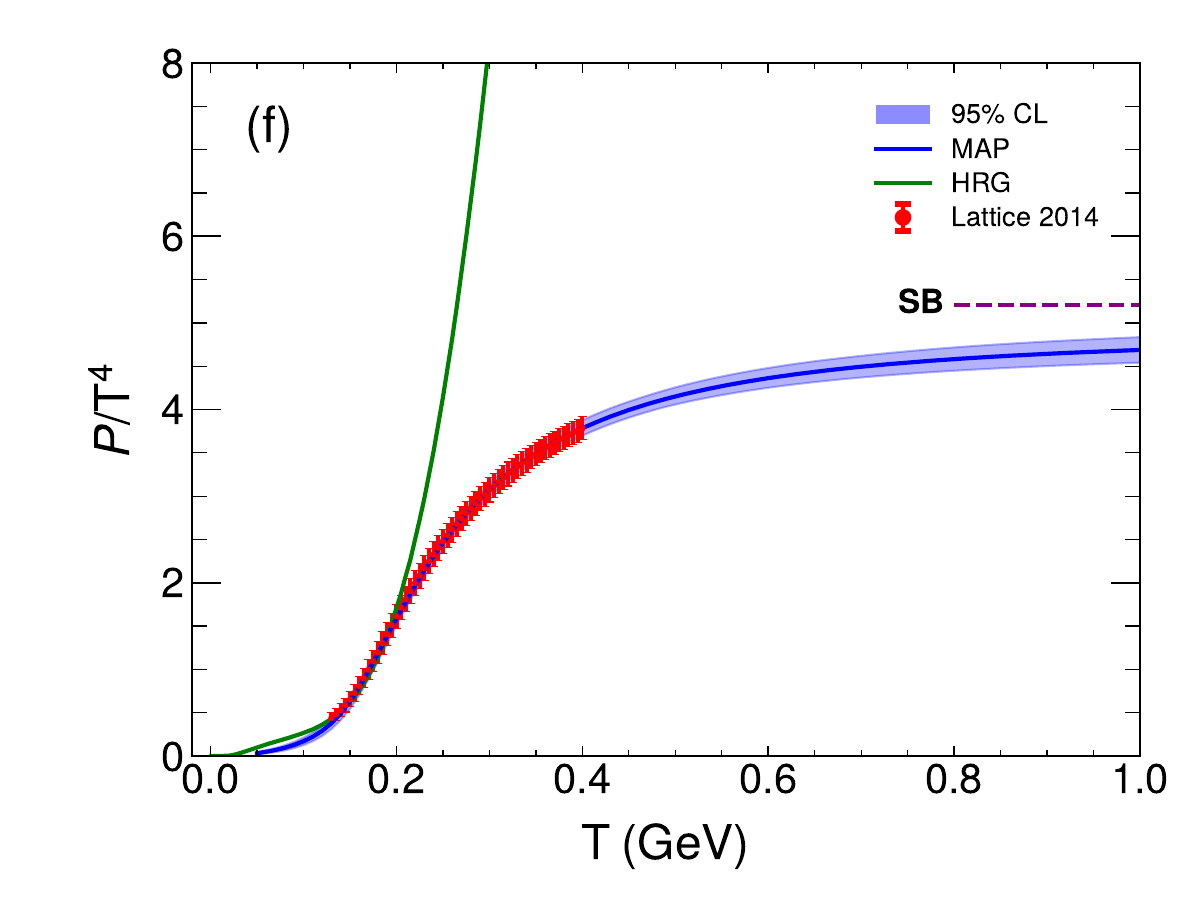}
    \end{minipage}             
    \begin{minipage}{0.40\textwidth}
        \includegraphics[width=\textwidth]{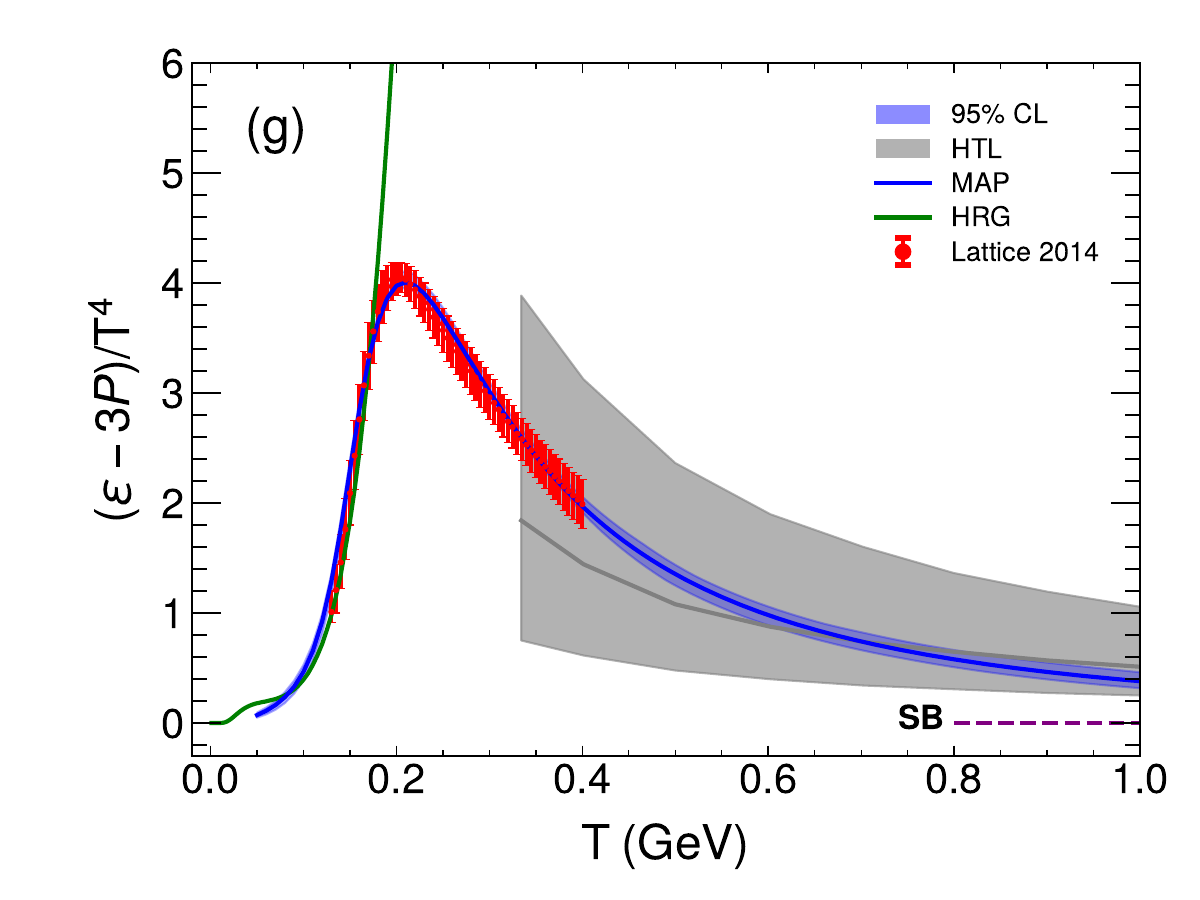}
    \end{minipage}
    \begin{minipage}{0.40\textwidth}
        \includegraphics[width=\textwidth]{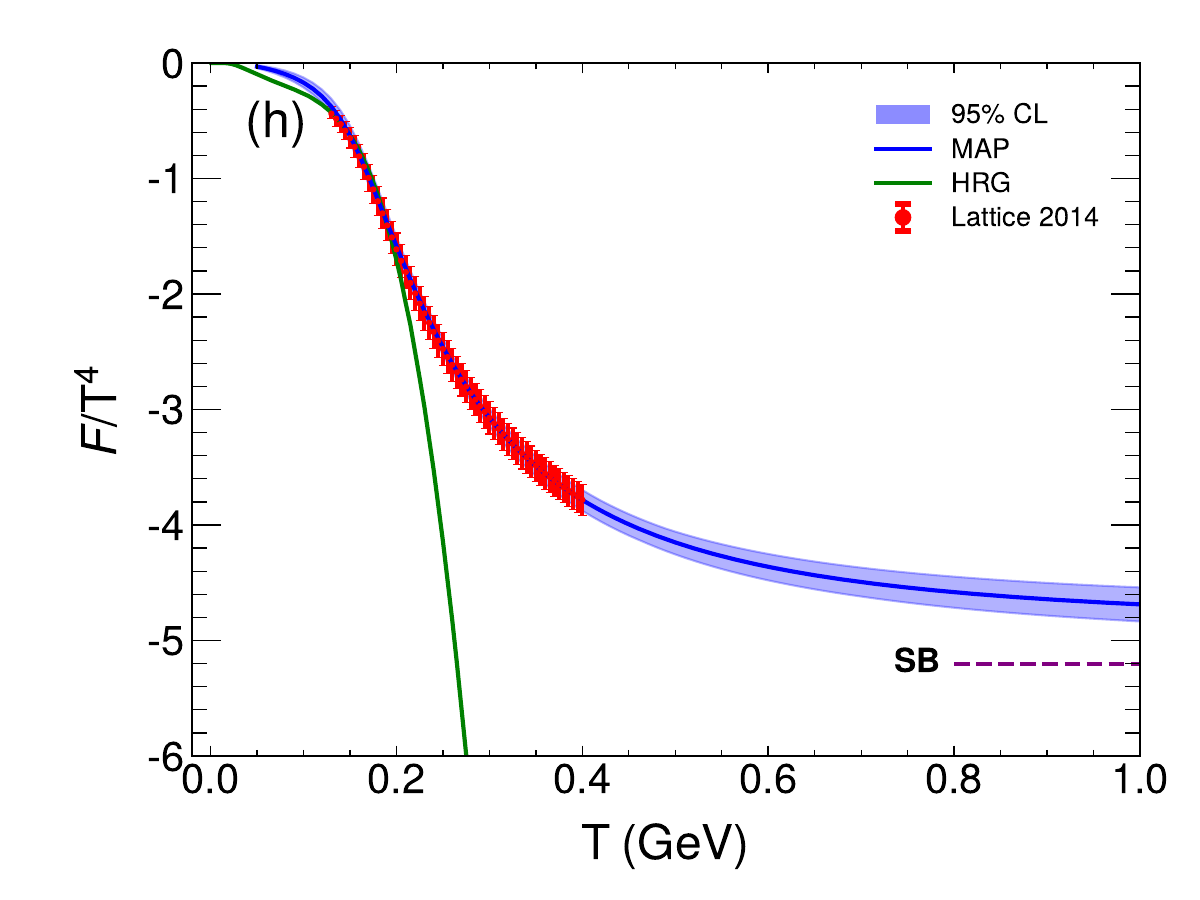}
    \end{minipage}
    \caption{The posterior parameter predictions of our model are shown in the figure (blue shaded bands: 95\% CL; solid curves: MAP), together with the LQCD data (red symbols with error bars), the HRG results (green solid line), and the HTL calculation (gray band). Panels (a)--(h) present the temperature dependence of the entropy, baryon number susceptibility, the squared speed of sound, the specific heat, energy, pressure, trace anomaly, and free energy, respectively. The LQCD results are taken from Refs.~\cite{HotQCD:2014kol,Bazavov:2017dus}, whereas the HRG and HTL curves are from Refs.~\cite{Vovchenko:2019pjl} and~\cite{Haque:2014rua}, respectively.}
    \label{fig4}
\end{figure*}

Figure~\ref{fig4} vividly illustrates the posterior outcomes of our model, where the blue shaded band delineates the 95\% CL, the solid blue curve represents the MAP estimate, the red symbols with error bars correspond to the LQCD data, and the green solid line together with the gray band depict the HRG and HTL predictions, respectively; SB marks the Stefan--Boltzmann ideal-gas limit.
In the intermediate temperature range ($150$ MeV $< T < 400$ MeV), all thermodynamic observables predicted by our model agree remarkably well with the LQCD data within the statistical uncertainties, demonstrating that the Bayesian calibration effectively captures the LQCD information. As the temperature rises further to $T > 400$ MeV, the squared speed of sound $C_s^2$ exhibits a smooth increase and gradually approaches the conformal limit of $1/3$~\cite{Borsanyi:2010cj}, this indicates that the restoration of conformal symmetry at elevated temperatures is properly accounted for by our model. In the low-temperature regime ($100$ MeV $< T < 150$ MeV), the calibrated holographic model yields $S/T^3$, $C_{V}/T^3$, $\epsilon/T^4$, and $(\epsilon-3P)/T^4$ that closely coincide with the HRG predictions, implying that the model naturally reproduces the thermodynamic features dominated by the hadron resonance gas in this region. Entering the high-temperature domain ($T > 400$ MeV), the trace anomaly $(\epsilon-3P)/T^4$ computed by our model falls within the HTL uncertainty band, which strongly confirms that our bayesian holographic QCD model correctly recovers the asymptotic perturbative QCD behavior.Nevertheless, over the examined temperature range, all quantities approach but stay below the SB limit, indicating that the system remains somewhat removed from a completely ideal gas regime. It is worth emphasizing that the LQCD evidence used in the Bayesian inference covers only the limited temperature window of $130$ MeV $< T < 400$ MeV. Overall, figure~\ref{fig4} clearly demonstrates that the calibrated holographic model not only precisely absorbs the nonperturbative information encoded in the LQCD equation of state, but also exhibits a robust extrapolation capability: it extends smoothly to both low and high temperatures, seamlessly connecting to the expected results of HRG and HTL.

Notably, the specific heat $C_{V}/T^{3}$ and the free energy $F/T^{4}$ both exhibit smooth and monotonic temperature dependence over the entire temperature range, without any jumps or singularities. This smooth behavior without any discontinuities is consistent with the crossover nature of the QCD transition at physical quark masses.

\subsection{Thermodynamics at finite Chemical potential}\label{sec4-2}

In this subsection, we employ the posterior parameters, constrained by LQCD results at zero chemical potential through bayesian inference, to further compute thermodynamic quantities at finite chemical potential and compare them with the corresponding LQCD results. Since the model parameters are entirely determined by observational data at zero chemical potential, the resulting calculations at finite chemical potential represent genuine theoretical predictions that do not rely on any additional parameter tuning. This therefore provides an independent and stringent test of the model's generalization capability and extrapolation reliability.

\begin{figure}[tbp!]
    \centering
    \includegraphics[width=0.43\textwidth]{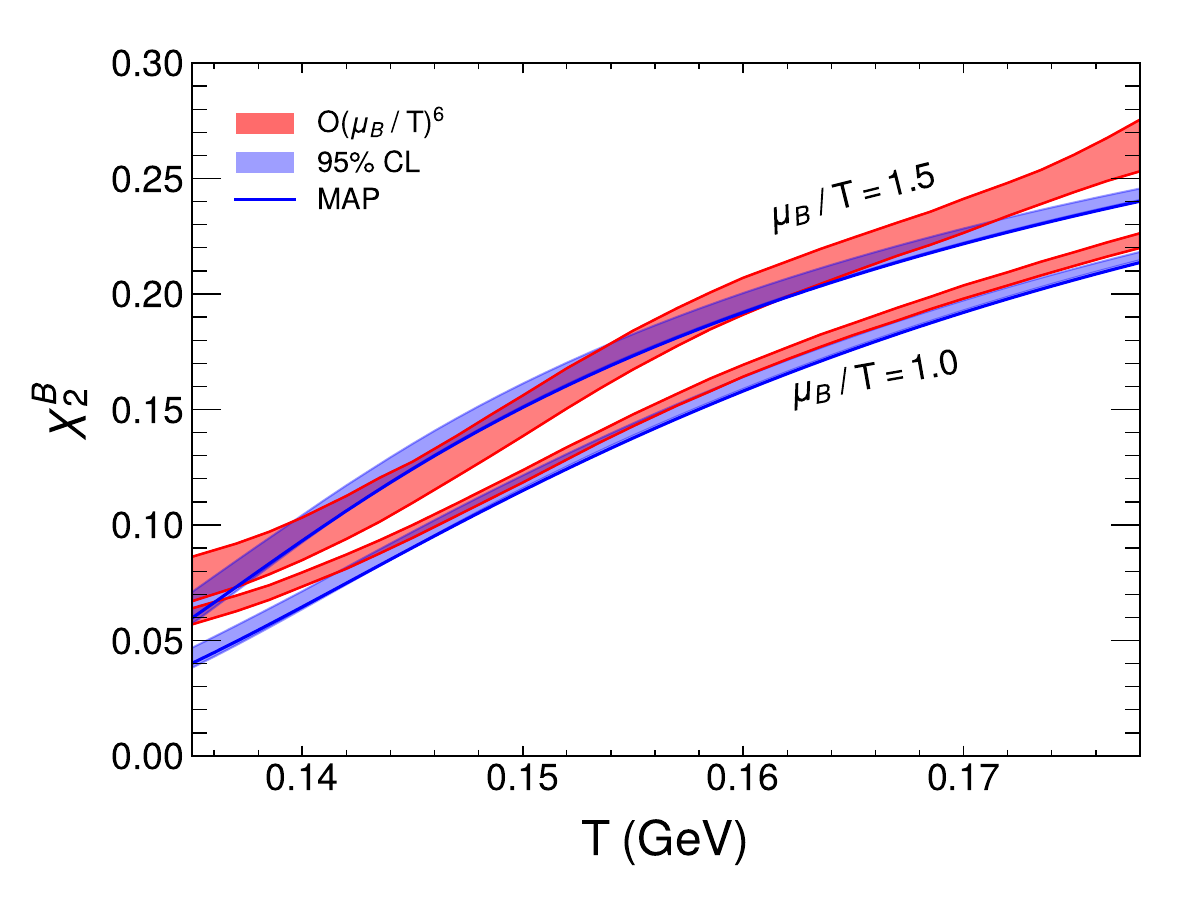}
    \caption{We compare the predictions of (2+1)-flavor LQCD and the bayesian holographic QCD model for the temperature dependence of the second-order baryon number susceptibility $\chi_2^B$ at finite chemical potential. Two representative values of the chemical potential are selected for the comparison: $\mu_B/T = 1.0$ and $1.5$. The red bands represent the LQCD results~\cite{Bollweg:2022rps}, while the blue bands and the blue solid lines correspond to the posterior 95\%~CL and MAP results of the bayesian holographic QCD model, respectively.}
    \label{fig5}
\end{figure}

Figure~\ref{fig5} presents the temperature dependence of the second-order baryon number susceptibility $\chi_2^B$ at two representative values of the chemical potential, $\mu_B/T = 1.0$ and $1.5$. The blue shaded band indicates the 95\% CL obtained from the bayesian holographic QCD model, while the solid blue curve corresponds to the MAP estimate. For comparison, the LQCD results, shown as a red error band, are derived from a Taylor expansion truncated at $\mathcal{O}((\mu_B/T)^6)$ and are taken from Ref.~\cite{Bollweg:2022rps}. It can be clearly seen that the bayesian holographic QCD model not only remains consistent with the LQCD Taylor expansion results within uncertainties over the entire temperature range, but also successfully reproduces the characteristic overall enhancement of $\chi_2^B$ with increasing chemical potential. This rising trend reflects the physical picture that baryon number fluctuations become more pronounced as the chemical potential grows, and it further demonstrates that the holographic model, calibrated solely with LQCD data at zero chemical potential, can be reliably extrapolated to the finite chemical potential regime, thereby providing reasonable theoretical predictions for baryon-sensitive thermodynamic quantities.

\begin{figure*}[tbp!]
    \centering
    \begin{minipage}{0.45\textwidth}
        \includegraphics[width=\textwidth]{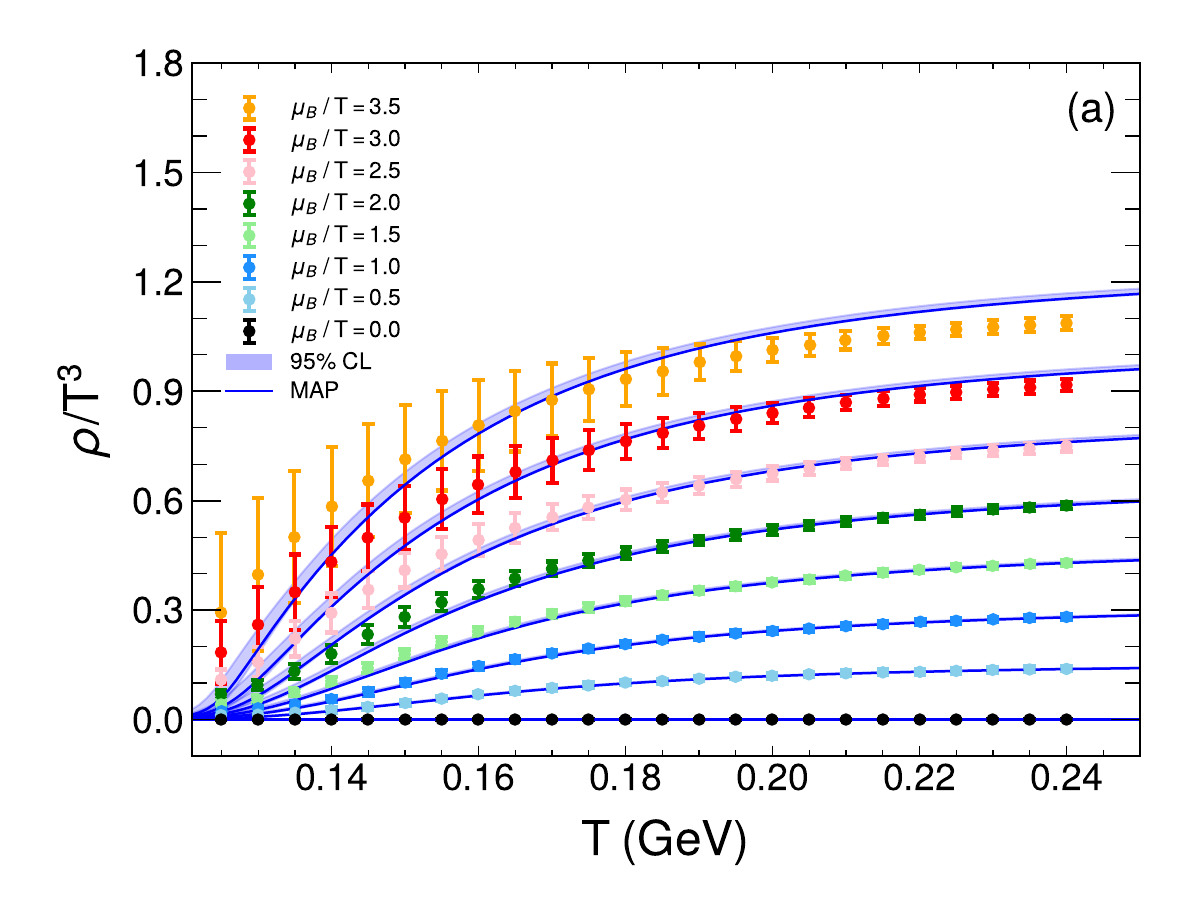}
    \end{minipage}
    \begin{minipage}{0.45\textwidth}
        \includegraphics[width=\textwidth]{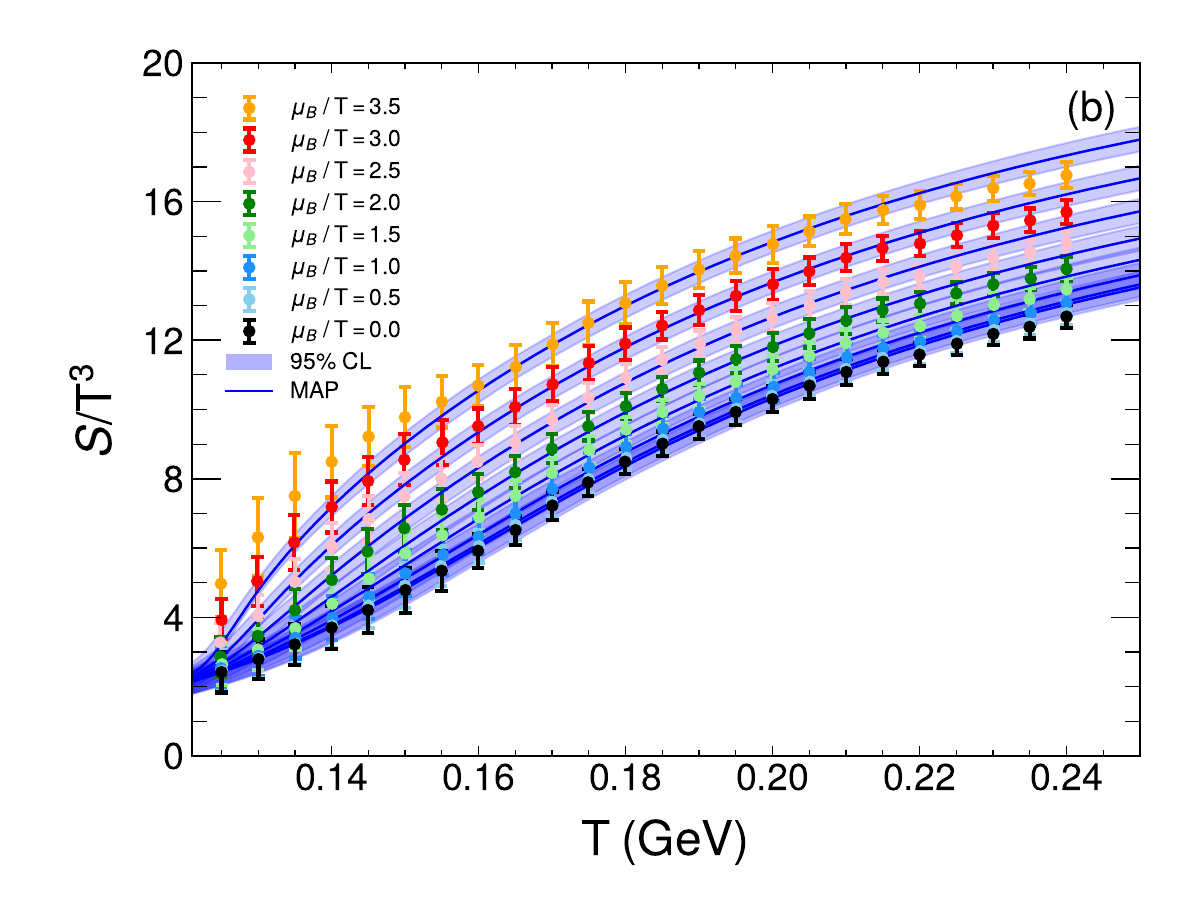}
    \end{minipage}
    \begin{minipage}{0.45\textwidth}
        \includegraphics[width=\textwidth]{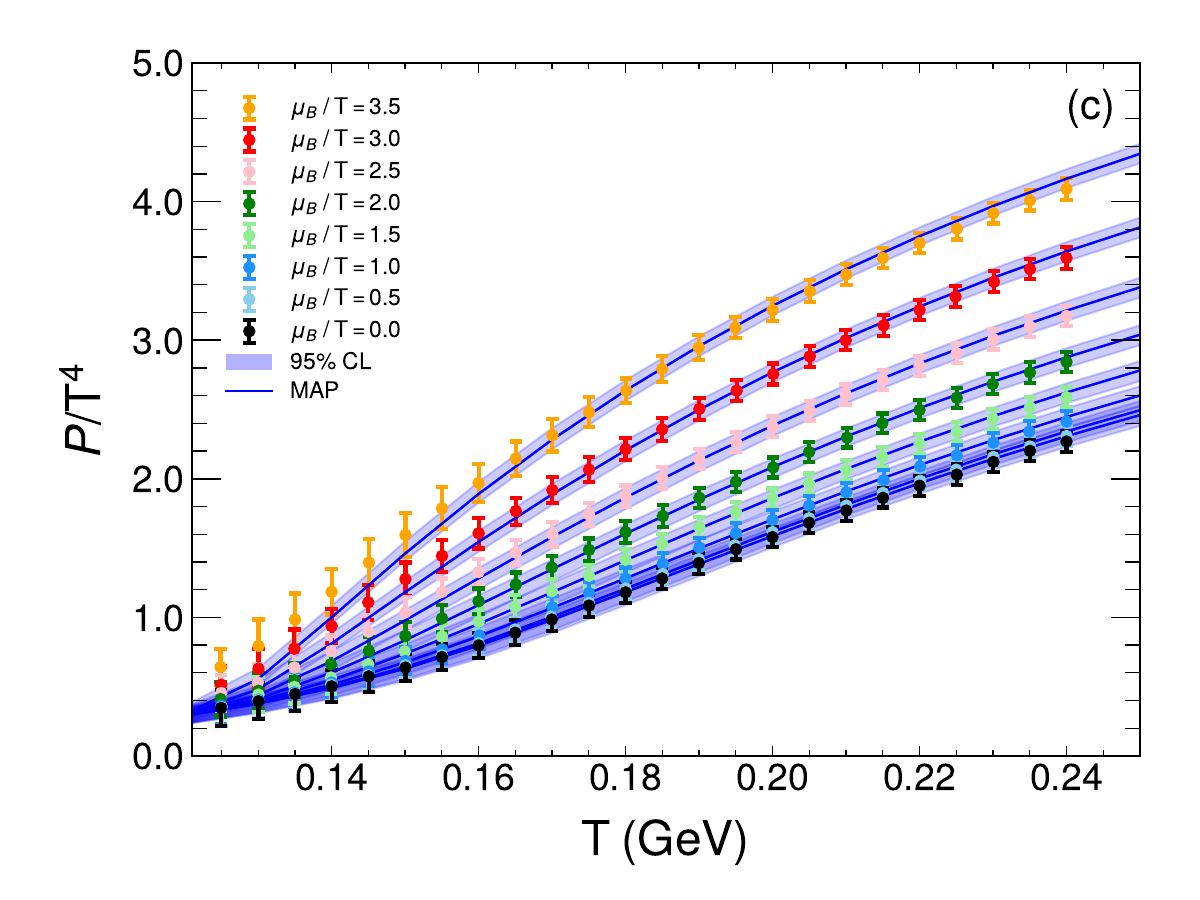}
    \end{minipage}
    \begin{minipage}{0.45\textwidth}
        \includegraphics[width=\textwidth]{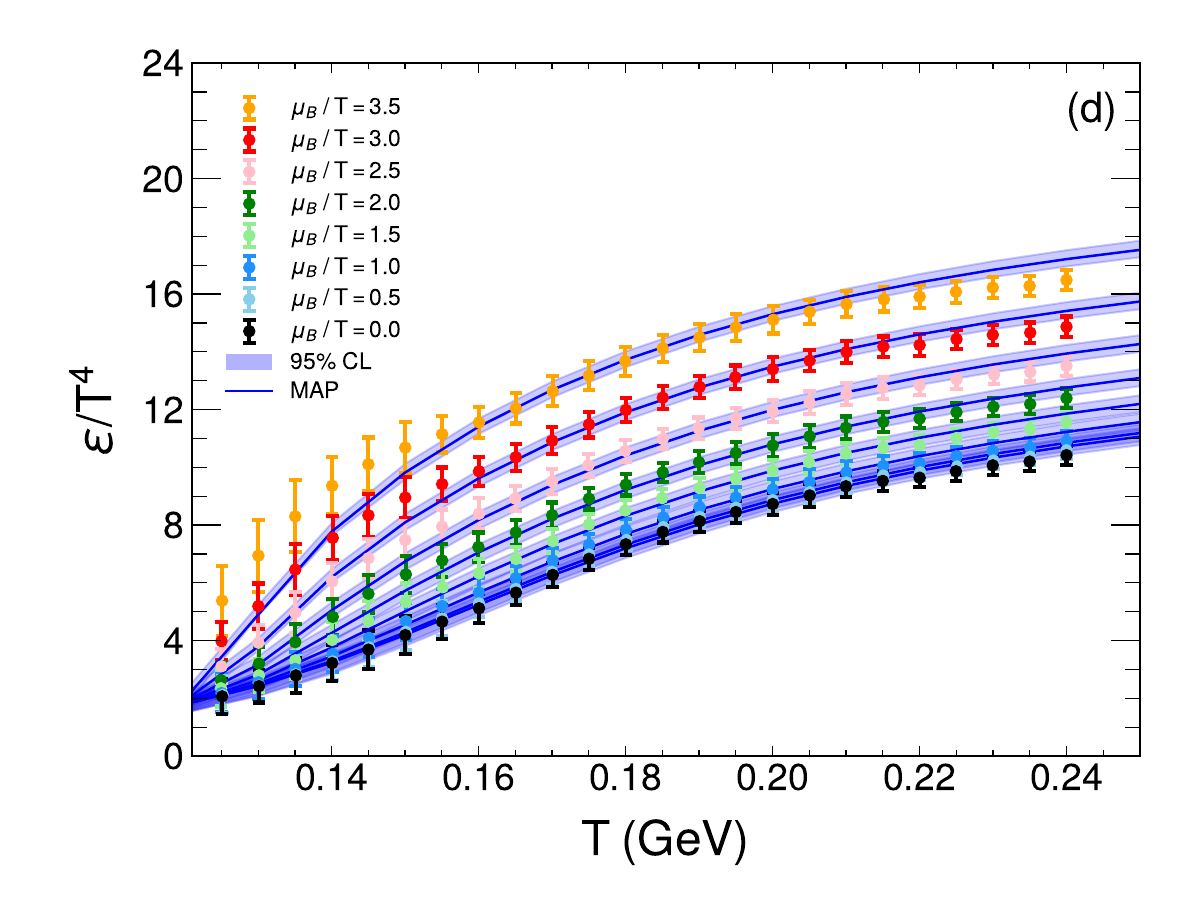}
    \end{minipage}
    \caption{We compare the predictions of (2+1)-flavor LQCD and the holographic model for the temperature dependence of several thermodynamic quantities at finite chemical potential, with $\mu_B/T$ ranging from $0$ to $3.5$. The individual panels display: (a) baryon number density $\rho/T^3$, (b) entropy $S/T^3$, (c) pressure $P/T^4$, and (d) energy density $\epsilon/T^4$. The data points with error bars represent the LQCD results~\cite{Borsanyi:2021sxv}, while the blue bands and blue solid lines correspond to the posterior 95\%~CL and MAP results of our bayesian holographic QCD model, respectively.}
    \label{fig6}
\end{figure*}

We present extrapolated results for several thermodynamic quantities at finite real chemical potential. The panels of figure~\ref{fig6} display the baryon density, entropy, pressure, and energy density over the range $\mu_{B}/T=0$--$3.5$.

Figure~\ref{fig6}(a) displays the temperature dependence of the baryon number density $\rho/T^3$ at finite baryon chemical potential $\mu_B$ for several representative values of $\mu_B/T$. We compare the predictions of the bayesian holographic QCD model with LQCD data~\cite{Borsanyi:2021sxv}. Overall, good agreement is observed, except for a slight overestimation by our model relative to the LQCD data within the temperature window $0.20$--$0.24$~GeV at $\mu_B/T = 3.0$ and $3.5$, and a slight underestimation at low temperatures for $\mu_B/T = 2.0$--$3.5$. In general, the bayesian holographic QCD model successfully captures the characteristic increase in baryon density with increasing chemical potential. 

Figure~\ref{fig6}(b) displays the temperature dependence of the entropy $S/T^3$ at finite baryon chemical potential $\mu_B$ for several representative values of $\mu_B/T$. Overall, the results of our Bayesian holographic QCD model are in good agreement with the LQCD data, with only a slight underestimation at low temperatures for $\mu_B/T = 2.5$--$3.0$. Furthermore, the results show that $S/T^3$ increases with increasing $\mu_B/T$. The increase of $S/T^3$ with increasing $\mu_B/T$ can be understood from the contribution of finite baryon chemical potential to the thermodynamic properties of the system. As $\mu_B/T$ grows, the net baryon number density in the system increases, and more baryonic degrees of freedom are excited, leading to an enhancement of the entropy. The fact that the holographic model naturally captures this trend indicates that its description of the thermodynamic properties of QCD matter at finite density is both reasonable and self-consistent.

Figure~\ref{fig6}(c) displays the temperature dependence of the normalized pressure $p/T^4$ at finite baryon chemical potential $\mu_B$ for several representative values of $\mu_B/T$. The model calculations show that $p/T^4$ increases systematically with increasing $\mu_B/T$, while remaining in good agreement with the LQCD data over the entire temperature range. Our calculations show that $P/T^4$ increases with $\mu_B/T$ over the temperature range $T = 0$--$0.24$~GeV, and this enhancement becomes more pronounced in the higher temperature region. This behavior can be understood from the contribution of a finite baryon chemical potential to the thermodynamic properties of the system. As $\mu_B/T$ increases, the net baryon number density grows, and the finite chemical potential provides an additional contribution to the pressure through the equation of state. At low temperatures, the system resides in the hadron resonance gas phase, where baryonic degrees of freedom are not yet fully excited, and the dependence of $P/T^4$ on the chemical potential is relatively weak. As the temperature rises, the system passes through the crossover transition and approaches the quark-gluon plasma phase, where more and more baryonic degrees of freedom are liberated, making the chemical potential contribution to the pressure increasingly significant. Consequently, the increase of $P/T^4$ with $\mu_B/T$ becomes more prominent in the higher temperature region. The good agreement between our holographic model results and the LQCD data over the entire temperature range indicates that the model provides a reasonable description of the thermodynamic behavior of QCD matter at finite density.

Figure~\ref{fig6}(d) displays the temperature dependence of the normalized energy density $\epsilon/T^4$ at finite baryon chemical potential $\mu_B$ for several representative values of $\mu_B/T$. The model calculations show that $\epsilon/T^4$ increases with increasing $\mu_B/T$. Apart from minor deviations in the low-temperature region, the predictions of our bayesian holographic QCD model for $\epsilon/T^4$ are in good agreement with the LQCD data over the entire temperature range.
The model calculations show that $\epsilon/T^4$ increases with increasing $\mu_B/T$. This behavior can be understood from the contribution of a finite baryon chemical potential to the energy density of the system. As $\mu_B/T$ increases, the net baryon number density grows, and the finite chemical potential introduces an additional energy contribution through the equation of state. In the high-temperature region, the system approaches the quark-gluon plasma phase, where baryonic degrees of freedom are fully excited, and the enhancement of the energy density due to the chemical potential becomes particularly pronounced. Consequently, $\epsilon/T^4$ increases monotonically with $\mu_B/T$, consistent with the behavior of the pressure and entropy density, reflecting the sensitivity of the equation of state of QCD matter at finite density to the chemical potential.

\begin{figure*}[tbp!]
    \centering
    \begin{minipage}{0.45\textwidth}
        \includegraphics[width=\textwidth]{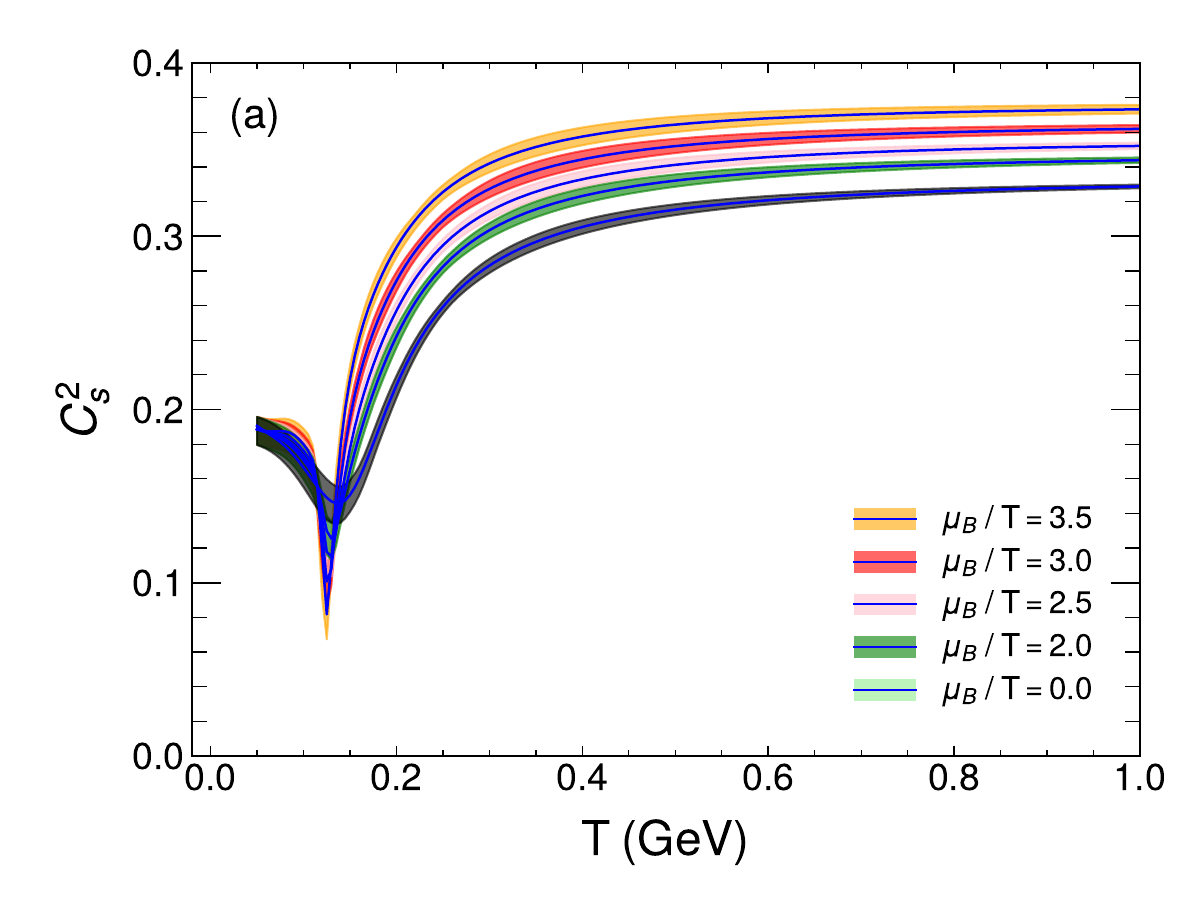}
    \end{minipage}
    \begin{minipage}{0.45\textwidth}
        \includegraphics[width=\textwidth]{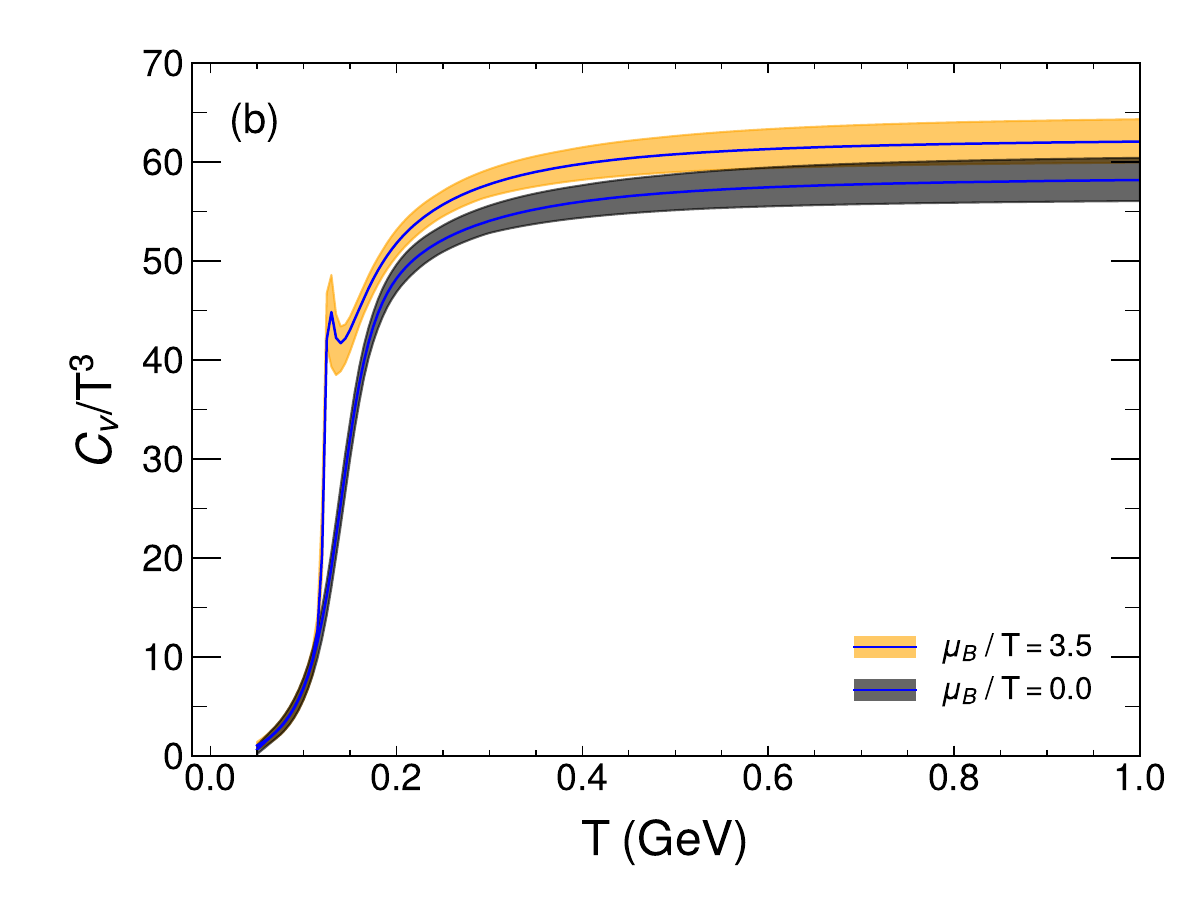}
    \end{minipage}
    \caption{We have calculated the temperature dependence of the speed of sound squared $C_s^2$ and the specific heat $C_V/T^3$ at finite chemical potential using the holographic model, with $\mu_B/T$ ranging from $0$ to $3.5$. Panel (a) presents the results for $C_s^2$, and panel (b) shows those for $C_V/T^3$. The colored bands represent the 95\%~CL results, while the blue solid lines correspond to the MAP results.}
    \label{fig7}
\end{figure*}

As shown in figure~\ref{fig7}, we have also calculated the temperature dependence of $C_s$ and $C_V$ at finite baryon chemical potential $\mu_B$ for several representative values of $\mu_B/T$. The results for $C_{s}^{2}$ are presented in figure~\ref{fig7}(a), and those for $C_{V}/T^{3}$ are shown in figure~\ref{fig7}(b).
The results show that both $C_{s}^{2}$ and $C_{V}/T^{3}$ increase with increasing $\mu_B/T$.
This behavior can be understood from the contribution of a finite baryon chemical potential to the thermodynamic properties of the system. As $\mu_B/T$ increases, the net baryon number density grows, leading to a simultaneous increase in both the energy density and the pressure. Resulting in a larger $C_{s}^{2}$. At the same time, the introduction of a finite chemical potential requires the system to absorb more energy to excite additional baryonic degrees of freedom as the temperature rises, which manifests as an enhancement of $C_{V}/T^{3}$. The increasing trends of these two observables are mutually consistent and indicate that the thermodynamic response of the system becomes more pronounced with increasing chemical potential.

\section{QCD Phase diagram in the Bayesian Holographic model} \label{sec5}

\begin{figure*}[tbp!]
    \centering
    \begin{minipage}{0.44\textwidth}
        \includegraphics[width=\textwidth]{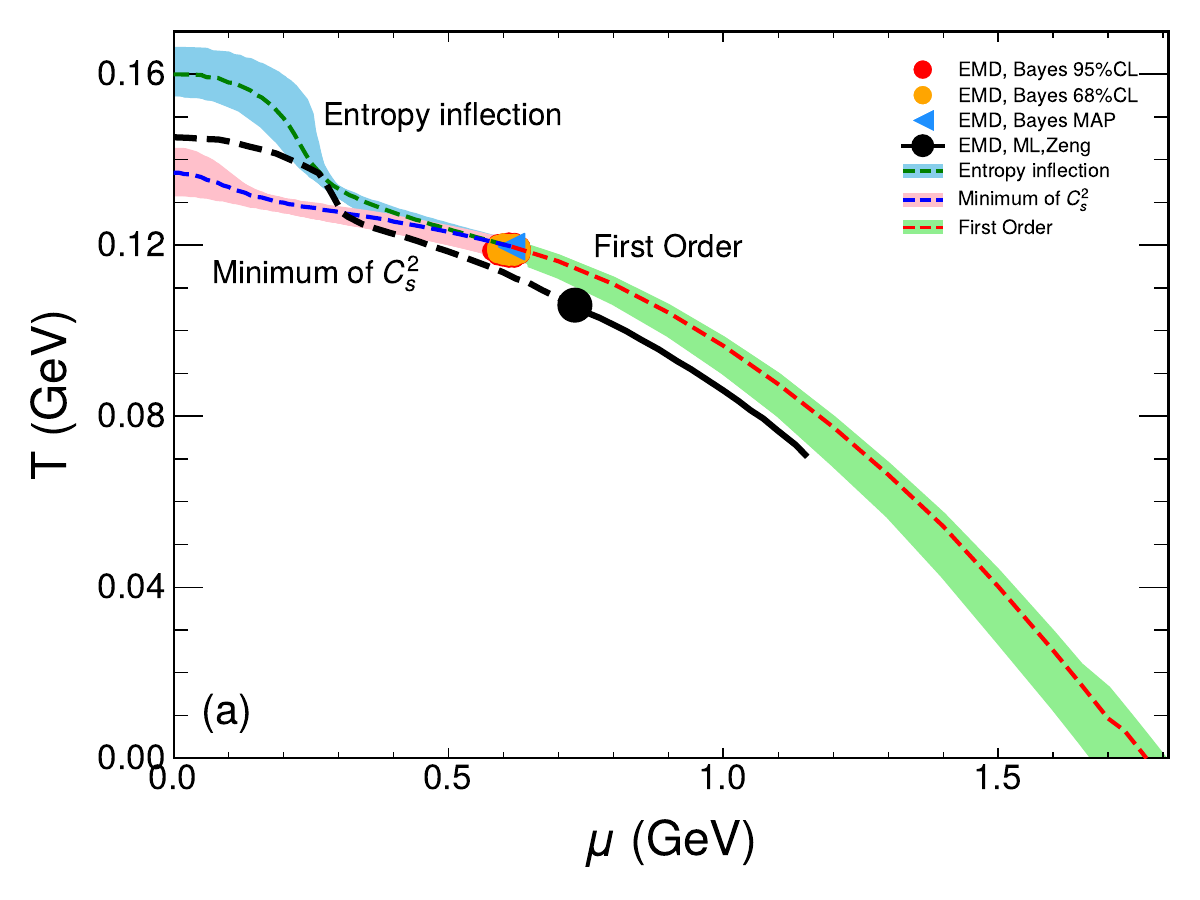}
    \end{minipage}
    \begin{minipage}{0.46\textwidth}
        \includegraphics[width=\textwidth]{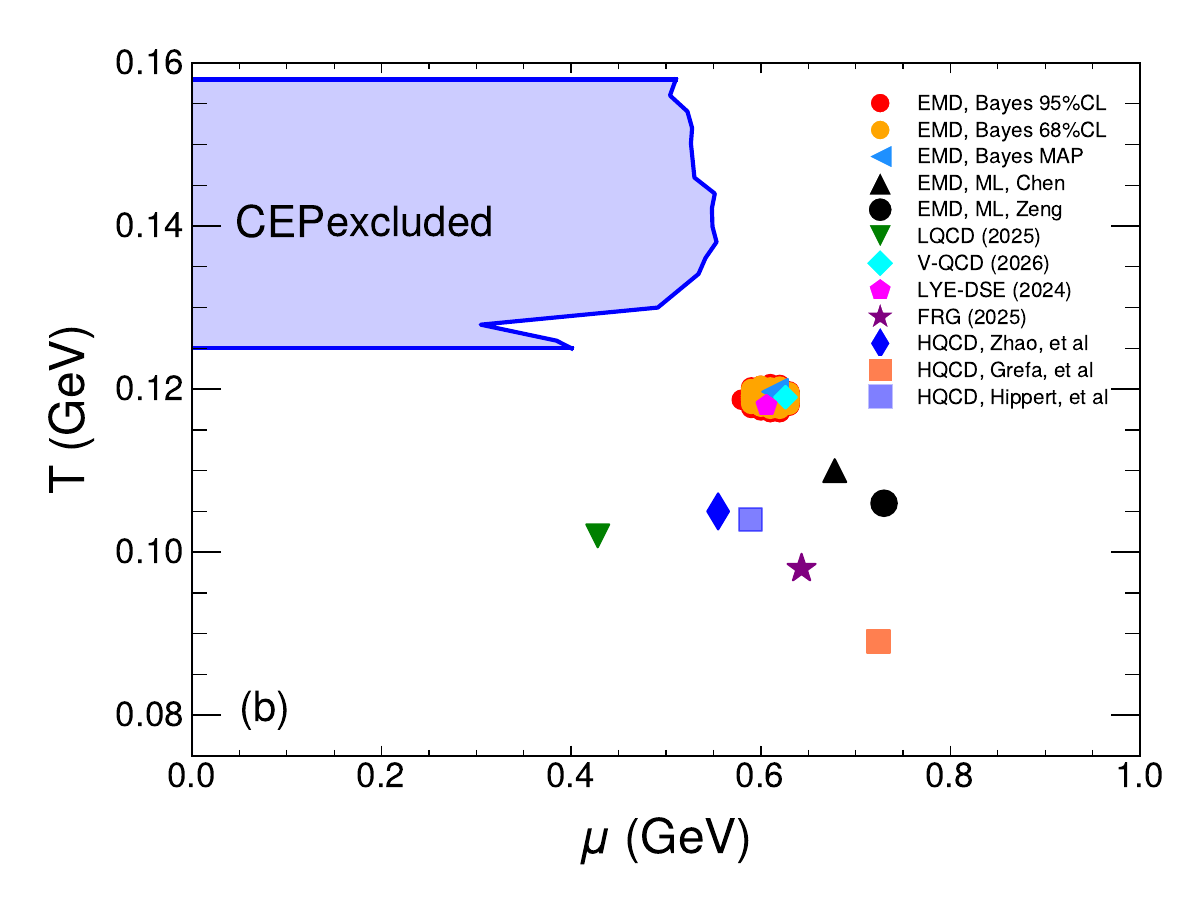}
    \end{minipage}
    \caption{(a) QCD phase diagram in the $T$--$\mu$ plane, For the crossover transition, the results based on the entropy criterion are indicated by the skyblue bands (95\% CL) and the green dashed lines (MAP), while those based on the speed-of-sound criterion are shown by the pink bands (95\% CL) and the blue dashed lines (MAP). For the first-order phase transition, the results obtained from the free-energy criterion are represented by the lightgreen bands (95\% CL) and the red dashed lines (MAP). As for the CEP, the red and orange dotted regions correspond to the 95\% CL and 68\% CL results, respectively. For comparison, the black dashed lines, black dots, and black solid lines denote the crossover transition, the CEP, and the first-order phase transition, respectively, obtained from other EMD model combined with machine learning~\cite{Zeng:2025tcz}. (b) Comparison of the CEP prediction from our model with those of other theoretical models, the red dotted region, the orange dotted region, and the dodgerblue left-pointing triangle represent our results at the 95\% CL, the 68\% CL, and the MAP estimate, respectively. For comparison, predictions from other theoretical models are also shown: the black upper triangles~\cite{Chen:2025goz} and black circles~\cite{Zeng:2025tcz} correspond to the machine learning assisted EMD model, the green lower triangle denotes the LQCD result~\cite{Clarke:2024ugt}, the cyan diamond indicates the holographic V-QCD prediction~\cite{Ecker:2025vnb}, the magenta pentagon represents the LYE-DSE result~\cite{Wan:2024xeu}, the pentagram is the FRG prediction~\cite{Fu:2023lcm}, and the blue diamond~\cite{Zhao:2022uxc}, orange square~\cite{Grefa:2021qvt}, and blue square~\cite{Hippert:2023bel} are predictions from holographic QCD models. The blue shaded region marks the area where no CEP exists~\cite{Borsanyi:2025dyp}.}
    \label{fig8}
\end{figure*}

In this section, we summarize the QCD phase diagram predicted by our model, together with the posterior distribution of the critical endpoint (CEP) obtained from bayesian inference, as shown in Figure.~\ref{fig8}. The phase diagram includes the crossover transition at small chemical potentials, the first-order phase transition line at larger $\mu_B$, and the location of the CEP constrained by the bayesian inference. The bayesian inference framework allows us to quantify the uncertainty in the CEP location through a statistically meaningful posterior distribution, based on a comprehensive analysis of the LQCD inputs.

Figure~\ref{fig8}(a) presents the QCD phase diagram predicted by our EMD model in the temperature--chemical potential ($T$--$\mu$) plane. The diagram clearly exhibits the characteristic features of the QCD phase structure: a crossover transition in the low-$\mu_B$ region, which gradually evolves into a first-order phase transition with increasing chemical potential.
For the crossover region, two different determination methods are adopted. The first method identifies the transition line by locating the inflection points of the gradient of $S/T^3$ with respect to temperature at fixed chemical potential, with the results displayed as the skyblue bands (95\% CL) and the green dashed lines (MAP). The second method determines the transition line by finding the minimum of the squared speed of sound as a function of temperature at fixed chemical potential, shown as the pink bands (95\%~CL) and the blue dashed lines (MAP).
The first-order phase transition line is obtained by determining, for each value of the chemical potential, the temperature at which the free energies of the two phases are equal~\cite{DeWolfe:2010he}, as indicated by the lightgreen bands (95\%~CL) and the red dashed lines (MAP).
The termination point of the first-order transition line is the CEP, the location of the CEP ~\cite{DeWolfe:2010he,He:2013qq,Yang:2014bqa,Yang:2015aia,Cai:2022omk} is determined by finding the maximum of the gradient of $S/T^3$ with respect to temperature at finite chemical potential. The red dotted region, the orange dotted region, and the dodgerblue left-pointing triangle denote our CEP predictions at the 95\%~CL, the 68\%~CL, and the MAP estimate, respectively. The corresponding numerical results are: $(T, \mu)_{68\% \mathrm{CL}} = (0.1176-0.1204,0.59-0.63)\mathrm{GeV}$, $(T, \mu)_{95\% \mathrm{CL}} = (0.1172-0.1206, 0.58-0.63)\mathrm{GeV}$, and $(T, \mu)_{\mathrm{MAP}} = (0.119, 0.615)\mathrm{GeV}$.
For comparison, the black dashed lines, black dots, and black solid lines represent the crossover transition, the CEP, and the first-order phase transition, respectively, obtained from other EMD models combined with machine learning techniques~\cite{Zeng:2025tcz}.
Similarly, we find that at \(T=0\), the range of \(\mu\) at the \(95\%\) CL is \(1.67\)--\(1.81\,\mathrm{GeV}\), while the value of \(\mu\) under the MAP estimate is \(1.77\,\mathrm{GeV}\).

Figure~\ref{fig8}(b) compares our model predictions (red and orange dotted region, dodgerblue left-pointing triangle) with results from other theoretical approaches. The black upper triangle at $(T, \mu) = (0.110, 0.678)$ GeV~\cite{Chen:2025goz} and the black circle at $(T, \mu) = (0.106, 0.730)$ GeV~\cite{Zeng:2025tcz} represent the predictions of the EMD model combined with machine learning techniques. 
Specifically, the black upper triangle is obtained by inputting the LQCD data on entropy density, baryon number susceptibility, and baryon number density into the EMD model and determining the model parameters with machine learning; the black circle is obtained by constructing a HoloNet neural network framework using machine learning methods, trained on the entropy density and baryon number susceptibility data from LQCD.
In contrast, our bayesian inference combined with the EMD model constrains the model parameters using the LQCD data on the entropy density, baryon number susceptibility, and the square of the speed of sound.
The green lower triangle at $(T, \mu) = (0.102, 0.428)$ GeV~\cite{Clarke:2024ugt} denotes the LQCD result, the cyan diamond at $(T, \mu) = (0.119, 0.626)$ GeV~\cite{Ecker:2025vnb} indicates the holographic V-QCD result, the magenta pentagon at $(T, \mu) = (0.118, 0.606)$ GeV~\cite{Wan:2024xeu} corresponds to the LYE-DSE result, the pentagram at $(T, \mu) = (0.098, 0.643)$ GeV~\cite{Fu:2023lcm} is the FRG result, and the blue diamond at $(T, \mu) = (0.105, 0.555)$ GeV~\cite{Zhao:2022uxc}, the orange square at $(T, \mu) = (0.089, 0.724)$ GeV~\cite{Grefa:2021qvt}, and the blue square at $(T, \mu) = (0.104, 0.589)$ GeV~\cite{Hippert:2023bel} are predictions from other holographic models. The results of the holographic V-QCD model and LYE-DSE fall within the range of our predictions. In particular, the CEP position predicted by the holographic V-QCD model, whose parameters have been fixed from neutron star observations, is consistent with the result obtained in this work, which is constrained by LQCD. This agreement between two independent lines of evidence further reinforces the validity of our model.

Finally, it should be noted that our current calculations and predictions rely on a specific model choice. Any modification to the holographic EMD model would consequently alter the Bayesian inference results, the posterior distributions, and the CEP predictions. The uncertainty in the final results originates from three sources: the errors in the input LQCD data, the errors introduced by the Gaussian emulator during the bayesian inference process, and the intrinsic uncertainty of the holographic EMD model itself. The first two have been taken into account in our calculations, while the model uncertainty arises from the assumed functional forms of $f(z)$ and $A(z)$.

\section{Conclusion} \label{sec6}

In this work, we have constructed a refined analytical bayesian holographic QCD model by systematically calibrating an EMD framework against (2+1)-flavor LQCD data through Bayesian inference. The model successfully describes the equation of state over a broad range of temperatures and baryon chemical potentials, exhibiting remarkable consistency with first-principles calculations. Compared with earlier studies~\cite{Chen:2024mmd, Chen:2024ckb}, the improved agreement with LQCD results at zero and finite chemical potential is achieved by introducing two additional parameters in the gauge kinetic function $f(z)$, and by performing the calibration using LQCD data for $S/T^3$, the second-order baryon number susceptibility $\chi_{2}^{B}$, and the squared speed of sound $C_{s}^{2}$ at zero chemical potential.

The central result of this study is the precise determination of the QCD phase diagram, as presented in figure~\ref{fig8}(a). Our model predicts a smooth crossover transition in the low-chemical-potential regime, which evolves into a first-order phase transition with increasing chemical potential and ultimately terminates at a critical endpoint (CEP). The inferred CEP location is $(T, \mu)_{68\%~\mathrm{CL}} = (0.1176\text{--}0.1204,\ 0.59\text{--}0.63)$~GeV and $(T, \mu)_{95\%~\mathrm{CL}} = (0.1172\text{--}0.1206,\ 0.58\text{--}0.63)$~GeV, with the maximum a posteriori estimate at $(T, \mu)_{\mathrm{MAP}} = (0.119,\ 0.615)$~GeV. Figure~\ref{fig8}(b) places this prediction within the broader theoretical landscape by comparing it with recent findings from other theoretical frameworks. The comparison shows that the results from the holographic V-QCD model and the LYE-DSE approach are in close agreement with our prediction.
Beyond the accurate description of the QCD phase structure, the bayesian holographic QCD model constructed in this work also demonstrates robust and comprehensive predictive power in describing both fundamental and derived thermodynamic quantities. At zero baryon chemical potential ($\mu_B=0$), the model not only faithfully reproduces the temperature dependence of key thermodynamic functions---including the entropy density, the second-order baryon number susceptibility, the squared speed of sound, the specific heat, the energy density, and the pressure---but also successfully captures derived observables such as the trace anomaly and the free energy, showing excellent agreement with LQCD data over the entire temperature range. More importantly, when extended to the finite chemical potential regime, the holographic predictions for the second-order baryon number susceptibility, the baryon number density, the entropy, the pressure, and the energy density remain broadly consistent with the LQCD results, further validating the reliability and extrapolation capability of this theoretical framework under finite-density conditions.

The comprehensive and remarkable consistency between our bayesian holographic QCD model and the (2+1)-flavor LQCD data, achieved under both zero and finite chemical potential conditions, provides compelling evidence that such bottom-up holographic approaches serve as a powerful and self-consistent theoretical tool for exploring the QCD phase diagram. More importantly, this framework, without relying on any prior assumptions, naturally extends the description into the finite-density regime that has long been inaccessible to standard LQCD simulations due to the fermion sign problem, thereby filling a critical gap left by first-principles calculations in this important region of the phase diagram. The accurate description of the equation of state, combined with the reliable constraint on the location of the critical endpoint (CEP), not only offers crucial theoretical guidance for ongoing heavy-ion collision experiments---such as the RHIC-STAR, LHC-ALICE, and other major international collaborative programs---but also lays a solid theoretical foundation for a deeper understanding of the properties of dense strongly interacting matter under the extreme conditions inside neutron stars, thus building a bridge connecting high-energy nuclear physics with astrophysical research.

\section{Acknowledgement}
This work is supported in part by the National Natural Science Foundation of China (NSFC) Grant Nos: 12405154, 12235016, 12221005, 12635009 and the Fundamental Research Funds for CUHK-Shenzhen university development fund under grant No.\ UDF01003041 and UDF03003041, Shenzhen Peacock fund under No.\ 2023TC0179.

\section{Appendix}

\begin{figure}[h]
    \centering
    \begin{minipage}{0.48\textwidth}
        \includegraphics[width=\textwidth]{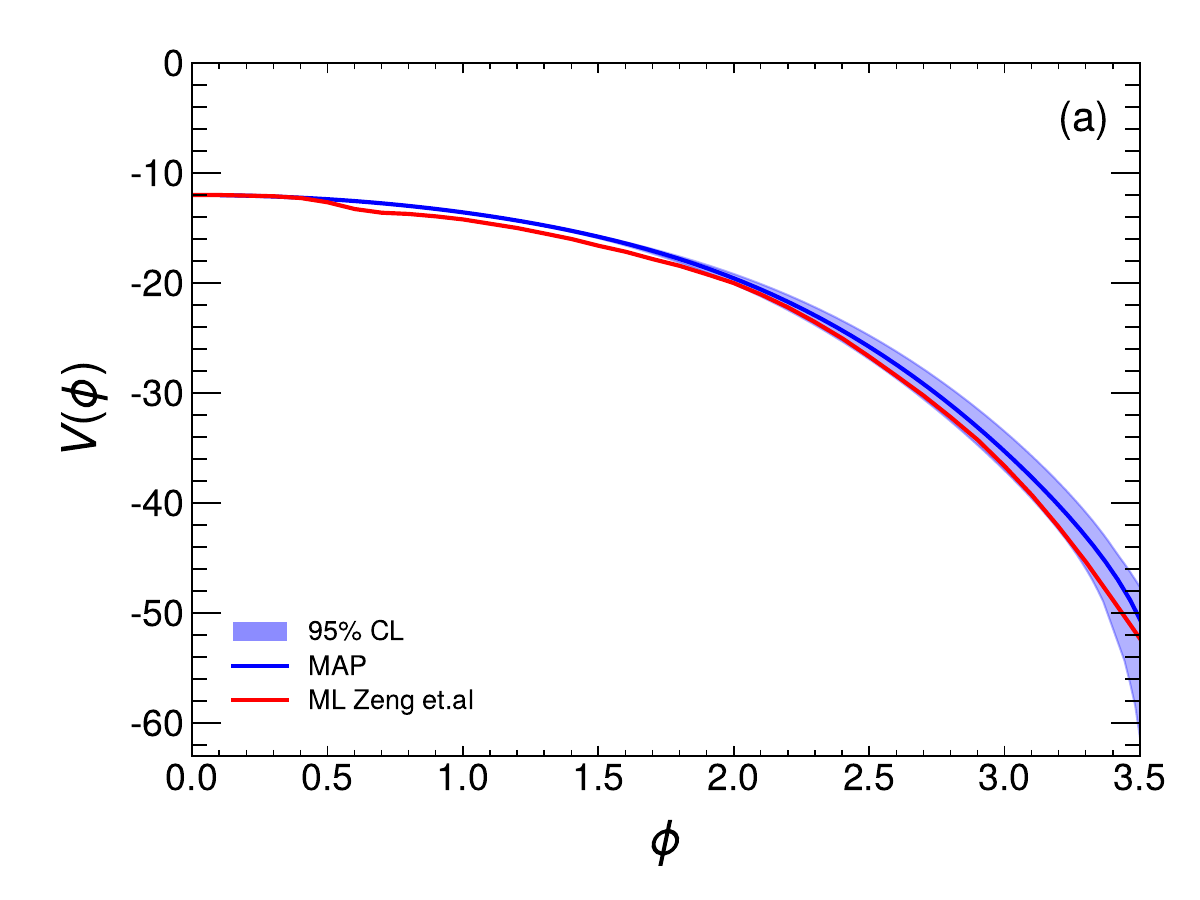}
    \end{minipage}
    \begin{minipage}{0.48\textwidth}
        \includegraphics[width=\textwidth]{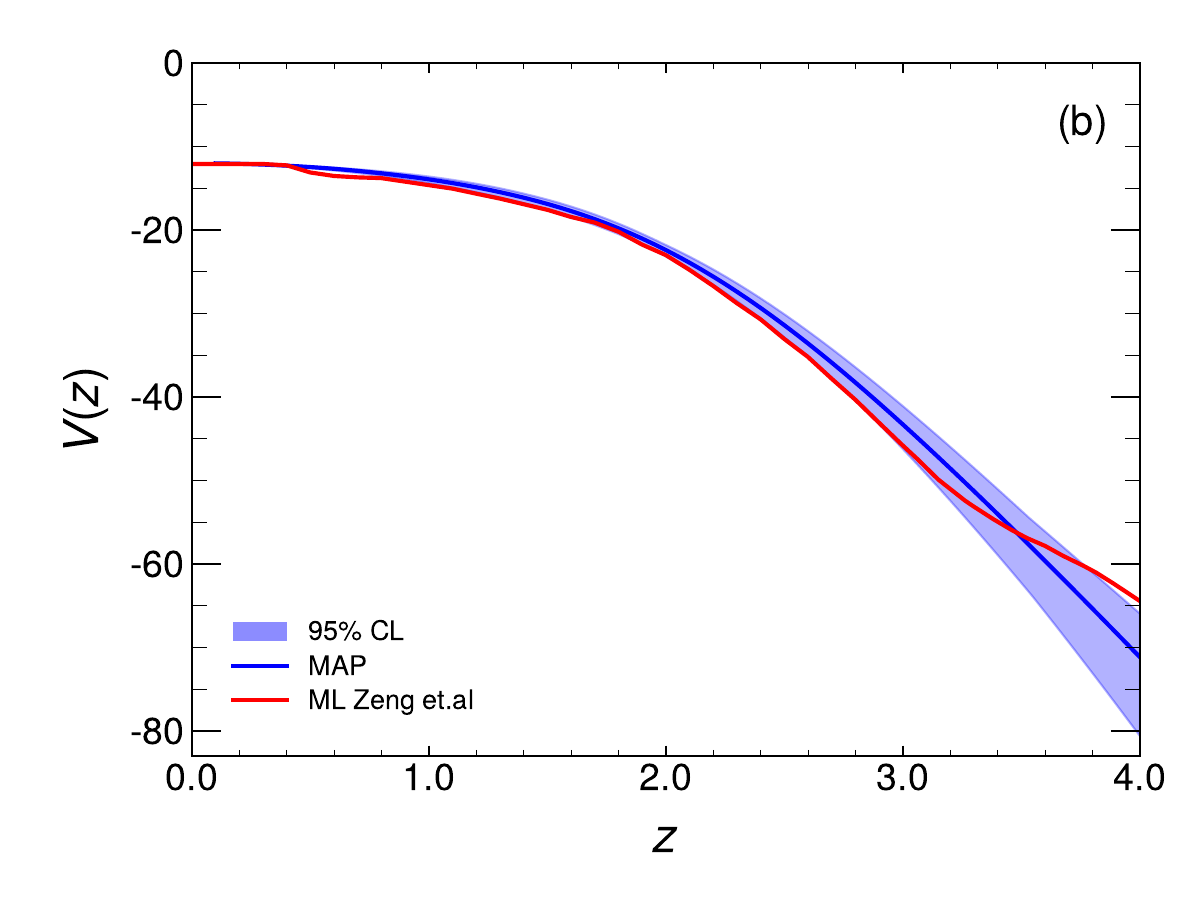}
    \end{minipage}
    \caption{We present the calculated results for \(V(\phi)\)--\(\phi\) and \(V(z)\)--\(z\) at the \(95\%\) CL (blue shaded band) and the MAP estimate (blue solid line), and compare them with the results from machine learning combined with the EMD model~\cite{Zeng:2025tcz}. Subplot (a) shows the relation between \(V(\phi)\) and \(\phi\), while subplot (b) shows the relation between \(V(z)\) and \(z\).}
    \label{fig9}
\end{figure}

Figure~\ref{fig9} presents the dilaton potential results obtained from bayesian analysis combined with the EMD model. Panel (a) shows \(V(\phi)\) as a function of the dilaton field \(\phi\), and panel (b) shows \(V(z)\) as a function of \(z\). The blue shaded band represents the parameter uncertainty at the 95\% CL, and the blue solid line denotes the MAP estimate. Meanwhile, the red solid line indicates the results from another machine learning combined with the EMD model ~\cite{Zeng:2025tcz} for comparison.
It can be observed that as \(\phi\) (or \(z\)) increases, the blue shaded band gradually widens, indicating that the model uncertainty increases in the larger \(\phi\) (or \(z\)) region. This is mainly because the numerical calculations are more sensitive to parameter variations in this region.
Notably, the overall trend of the red solid line is highly consistent with our MAP estimate. Specifically, in the regions \(\phi \lesssim 2.5\) and \(z \lesssim 2.5\), the red solid line is basically close to the blue solid line; in the larger \(\phi\) and \(z\) regions, although the red solid line slightly deviates from the MAP curve, it largely falls within our 95\% CL blue shaded band (with a slight deviation from the 95\% CL range near \(z \sim 4.0\)). This indicates that the results from other machine learning combined with the EMD model agree well with our bayesian analysis results within statistical uncertainties, further validating the reliability and robustness of our model calculations.

\bibliography{ref_temp}
\end{document}